\documentclass[preprint,12pt]{elsarticle}

\usepackage{amssymb}
\usepackage{amsmath}
\usepackage{amsthm}
\usepackage{siunitx}
\usepackage{float}

\usepackage{lineno}

\begin{document}

\begin{frontmatter}



\title{Experimentally validated process–microstructure–property relations of bainitic steels derived from phase-field simulations}


\author[a]{Dhanunjaya Kumar Nerella\corref{cor1}}
\author[a]{Muhammad Adil Ali}
\author[b]{Oguz Gulbay}
\author[a]{Oleg Shchyglo}
\author[a]{Ingo Steinbach}

\cortext[cor1]{Dhanunjaya Kumar Nerella}

\affiliation[a]{{Interdisciplinary Centre for Advanced Materials Simulation, Ruhr University Bochum},
            addressline={Universitaet str 150}, 
            city={Bochum},
            postcode={44801}, 
            state={NRW},
            country={Germany}}
            
\affiliation[b]{{Steel Institute, RWTH Aachen University},
            addressline={Intzestraße 1}, 
            city={Aachen},
            postcode={52072}, 
            state={NRW},
            country={Germany}}

\begin{abstract}
This study examines the impact of processing conditions, such as thermal processing, on the resulting microstructure and mechanical properties. Special emphasis is placed on microstructural features obtained from three-dimensional phase-field simulations, which provide detailed insights into bainite morphology, phase distribution and retained austenite content. These simulated microstructures are correlated with changes in yield strength under multiaxial load, as represented in the yield surface of the material. The results demonstrate that optimized processing routes can refine the microstructure, enhance mechanical properties and significantly alter the yield surface characteristics. These findings provide valuable insights for the design and application of bainitic steels, as well as for the development of predictive models linking process-structure-property relationships.
\end{abstract}

\begin{graphicalabstract}
\begin{figure}[H]
    \centering
    \includegraphics[width=1\textwidth]{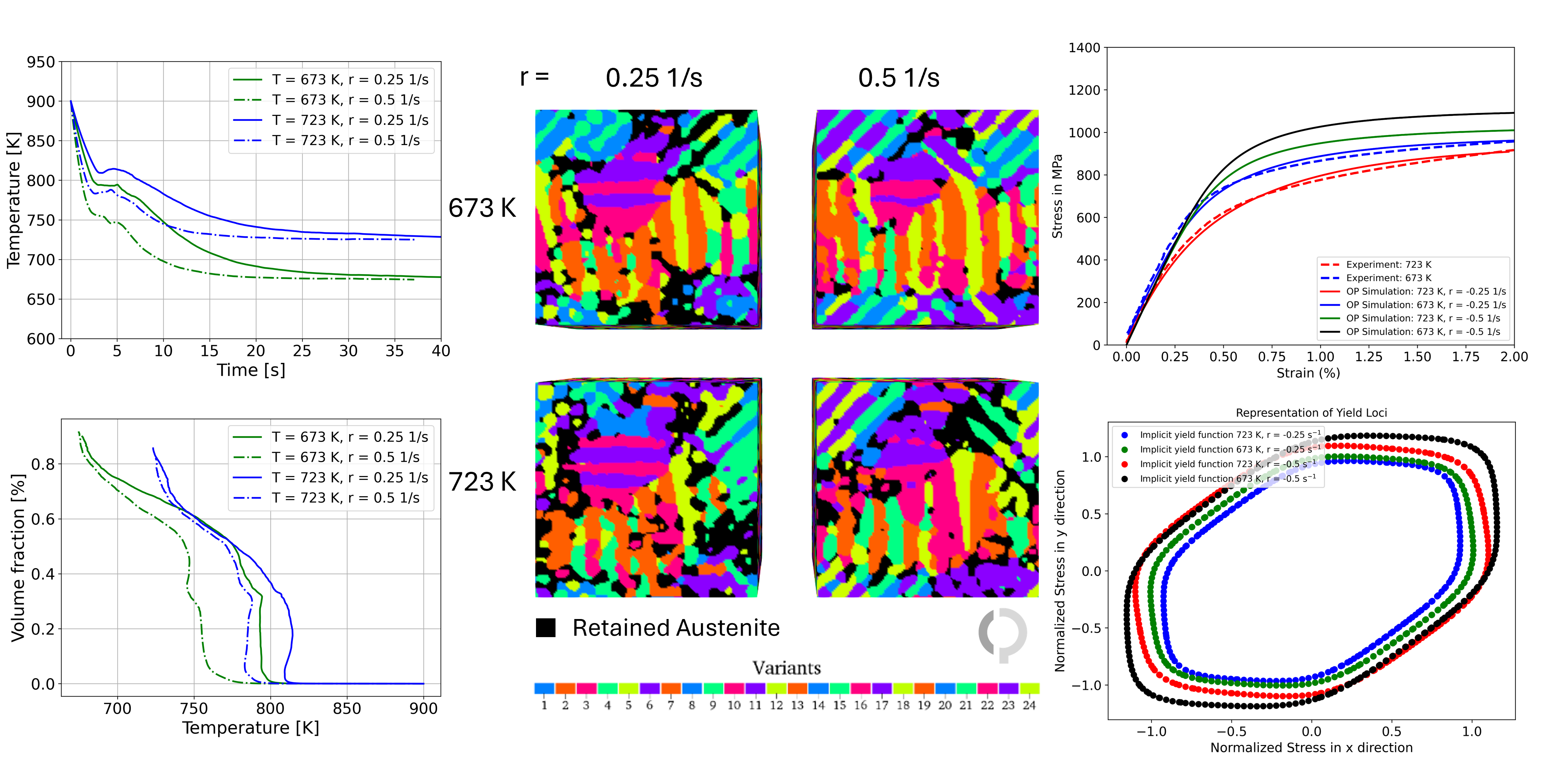}
\end{figure}
\end{graphicalabstract}

\begin{highlights}
\item Phase-field simulations quantify the effects of holding temperature and heat extraction rate on bainitic transformation kinetics, morphology and internal-stress development.
\item Lower holding temperature and higher heat extraction rate promote finer, more elongated bainitic ferrite, higher transformation-induced stresses and increased tensile strength.
\item Phenomenological yield-surface modelling reveals heat-treatment-dependent plastic anisotropy and links bainitic microstructure, internal stresses and anisotropic yielding.
\end{highlights}


\end{frontmatter}



\section{Introduction}

Bainitic steels are often chosen for their exceptional combination of high strength, fracture toughness and wear resistance, making them indispensable in demanding applications across the automotive, aerospace and heavy engineering sectors \cite{hajizad2019influence,fan2022effect,fielding2013bainite}. Their unique microstructural architecture enables a balance between tensile strength and ductility that is difficult to achieve in conventional steels \cite{gulbay2023influence}. Fully exploiting this potential, however, requires a quantitative understanding of the links between processing, microstructural evolution and mechanical response under complex loading conditions. Predictive capabilities in this domain are critical for the design of optimized components and the development of advanced manufacturing strategies.

Bainitic microstructure studies are carried out in \cite{gulbay2023influence,nerella2025automated}, where the influence of holding temperature on morphology have been presented. In \cite{nerella20262d}, 3D morphology of bainitic ferrite variants are studied for different external applied strains. In \cite{bakhtiari2009effect}, effect of bainite morphology on mechanical properties like yield strength, ultimate tensile strength and fatigue are studied. In \cite{zhao2019effect}, the influence of aspect ratio of bainite on deformation between ferrite and bainite is studied. Consistent with these studies, upper-bainite morphology carbide-free, interlath-cementite or intralath platelet is dictated by cooling/holding and tempering and governs crack-path scale, anisotropy. Appropriately heat treated bainite achieves higher toughness and better strength \cite{ohtani1990morphology}.

The phase-field method has emerged as a powerful computational approach for simulating the microstructural evolution of bainitic steels. By explicitly resolving phase transformations, grain morphology and interface motion, it captures the intricate formation pathways of bainite and the resulting spatial distribution of constituent phases \cite{steinbach2024highly,steinbach2013phase}. The Phase-Field (PF) simulations provide three-dimensional, time-resolved insights into phase fractions, morphology and retained austenite distribution, thereby establishing a physically informed basis for mechanical property prediction. In this study a coupled phase-field and phenomenological crystal plasticity framework is employed to study bainitic transformation.The formulation builds on finite strain phase-field descriptions of bainitic/martensite transformations \cite{Yeddu2012,Levitas2013,Salama2024} and on coupled models that incorporate the interaction between phase transformations and plastic deformation\cite{Levitas2015,Javanbakht2015}.The simulations also incorporate a recently developed finite-strain elasticity solver\cite{Shchyglo2024}.

Yet, the reliability of such predictions depends strongly on rigorous experimental validation. In this work, simulated microstructures are benchmarked against high-resolution observations from scanning electron microscopy (SEM) enabling quantitative comparison of bainitic plate thickness, phase fractions and morphology. Complementary experimental tensile tests provide the stress–strain response under uniaxial loading, allowing for direct validation and calibration of the simulation framework.

To extend the predictive capability beyond uniaxial loading, the anisotropic plastic behavior of bainitic steels is characterized using the Barlat91 yield criterion \cite{barlat91}. The Barlat91 phenomenological model is well-suited to capturing anisotropic yield surfaces in materials with anisotropic plasticity \cite{khalfallah2015influence,chaparro2008material}. Here, it is calibrated against experimental data to construct yield surfaces corresponding to different processing-induced microstructures, thereby linking microstructural variations to macroscopic plastic flow behavior under multiaxial stress states.

In addition to these advancements, understanding how processing variables, such as cooling rates, tempering conditions and alloy composition, influence the formation of bainitic microstructures is crucial for tailoring the material’s mechanical performance for specific applications \cite{zhao2025review,kaikkonen2023evaluation,soliman2008phase}. For instance, bainitic steels used in automotive and railway applications demand high tensile strength without compromising ductility and fracture toughness, while aerospace components require a balance of lightweight design and superior fatigue resistance~\cite{xiao2024design}. By developing a deeper understanding of the process–structure–property relationship, this study seeks to optimize bainitic steels for diverse engineering applications \cite{nanda2019third,wang2022novel,liang2014study}, paving the way for more reliable and efficient materials in high-performance systems.

By integrating phase-field simulations, experimental microstructural characterization, tensile testing and advanced yield surface modeling, this study establishes a comprehensive process–structure–property framework for bainitic steels. The results provide fundamental insight into the role of processing conditions in controlling microstructure and yield surface evolution, offering practical guidelines for the design of high-performance bainitic steels tailored to specific engineering applications.

\section{Experimental details}

An 80~kg steel ingot with a nominal composition of Fe-0.2C-1.5Si-2.5Mn (wt.\%) was produced in a laboratory-scale vacuum induction furnace. The ingot, with a cross-section of $140 \times 140~\si{\milli\meter\squared}$, was homogenized at \SI{1200}{\degreeCelsius} and subsequently forged down to billets with a cross-section of $60 \times 60~\si{\milli\meter\squared}$. A second homogenization treatment was then carried out for 5~h, followed by furnace cooling. The measured chemical composition is shown in Table~\ref{tab:composition}.

\begin{table}[htbp]
\centering

\begin{tabular}{ccccccccc}
\hline
C & Si & Mn & P & S & Cr & Mo & Al & Cu \\
\hline
0.19 & 1.48 & 2.38 & 0.003 & 0.003 & 0.04 & 0.01 & 0.003 & 0.02 \\
\hline
\end{tabular}

\caption{Chemical composition of the investigated steels in wt.-\%. Optical emission spectroscopy (OES) was used to determine the composition, and the carbon content was measured by combustion analysis.}
\label{tab:composition}
\end{table}

\begin{figure}[H]
    \centering
    \includegraphics[width=1\textwidth]{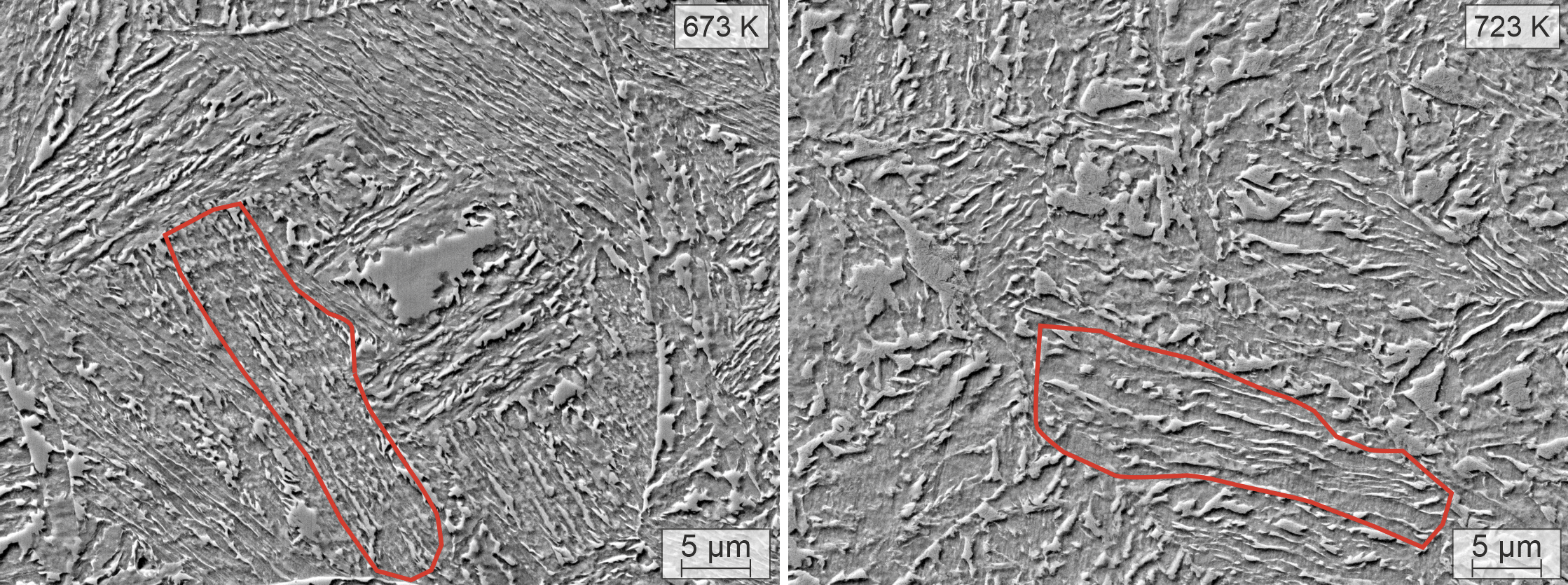}
    \caption{\label{fig:sem_image01}SEM images obtained via SE detector for microstructures held at 673 K and 723 K.}
\end{figure}

The specimens were austenitized in a salt bath at 60~K above the $A{\mathrm{c}_{3}}$ temperature for 300~s to obtain a fully austenitic microstructure. They were then rapidly transferred to a second salt bath and held isothermally at 673 K or 723 K for 45~min to achieve two different bainitic microstructure. After the isothermal holding, the specimens were quenched to room temperature. The heat treatments described above resulted in the microstructures shown in Figure~\ref{fig:sem_image01}. As intended, the bainitic microstructure were obtained in both conditions, consisting of a bainitic ferrite matrix. The bainitic ferrite formed at 673 K is noticeably finer compared to that formed at 723 K (highlighted in red color), as evident in the micrographs. Further description of microstructure is discussed in section~\ref{micro_expt}

\section{Phase-field Framework}

Phase-field modelling provides a thermodynamically consistent framework for simulating microstructural evolution in steels by describing interfaces implicitly through continuous field variables. In this approach, each phase or grain is represented by a phase-field parameter whose temporal evolution is governed by the minimization of a total free energy functional comprising chemical, interfacial and elastic contributions. The chemical free energy is derived from bulk thermodynamic descriptions of the relevant phases, while interfacial energies are introduced through gradient terms that enforce a finite interface thickness. Elastic effects arising from lattice misfit or transformation strains are incorporated via eigenstrain formulations, enabling the treatment of stresses and their influence on transformation kinetics and morphology. The resulting phase-field evolution equations naturally capture complex phenomena such as bainitic ferrite or martensite variant selection, impingement and anisotropic growth during solid-state phase transformations in steels under thermal or mechanical loading. 

The temporal evolution of the phase fields is governed by the multi phase-field kinetic equation describing microstructure evolution \cite{steinbach1999generalized,steinbach2006multi}:

\begin{equation}
\dot{\phi}_{\alpha}(\mathbf{x},t) =
-\frac{1}{N}
\sum_{\beta \neq \alpha}
M_{\alpha\beta}
\left(
\frac{\delta F}{\delta \phi_{\alpha}}
-
\frac{\delta F}{\delta \phi_{\beta}}
\right),
\end{equation}
where $M_{\alpha\beta}$ is the interface mobility between phases $\alpha$ and $\beta$. F is the total free energy and is expressed as

\begin{equation} 
\label{eq:free_energy}
F = \int_\Omega \left(
f_\text{chem}
+ f_\text{int}
+ f_\text{el}
\right),
\end{equation}
where $f_\text{chem}$ is the chemical free energy density, $f_\text{int}$ represents the interfacial contribution and $f_\text{el}$ accounts for the elastic strain energy.

The governing equations summarized above form the foundation of the present multi-phase-field framework for simulating microstructure evolution in polycrystalline materials undergoing bainitic transformations. In this approach, transformation proceeds through the coupled action of thermodynamic driving forces, interfacial and diffusion-controlled kinetics and mechanical interactions arising from transformation strains and elastic–plastic accommodation. The evolving microstructure is obtained by continuously minimizing the total free-energy functional.
The interfacial free-energy density $f_{int}$ in Eq.~\ref{eq:free_energy}  accounts for the energy within the boundaries separating different phases or grains:

\begin{equation}
f_{\mathrm{intf}} =
\sum_{\alpha=1}^{N}
\sum_{\beta>\alpha}^{N}
\frac{8\sigma_{\alpha\beta}}{\eta}
\left[
-\frac{\eta^{2}}{\pi^{2}}
\nabla \phi_{\alpha} \cdot \nabla \phi_{\beta}
+ \phi_{\alpha}\phi_{\beta}
\right],
\end{equation}
where $\phi_{\alpha}$ and $\phi_{\beta}$ are phase-field variables, $\sigma_{\alpha\beta}$ is the interface energy between phases or grains $\alpha$ and $\beta$ and $\eta$ denotes the diffuse interface width.

The chemical free-energy density, $f_{chem}$, captures the contribution of temperature and composition to the system’s free energy and thus provides the thermodynamic driving force for the phase transformation \cite{heo2014phase,moelans2008introduction}:

\begin{equation}
f_{\mathrm{chem}} =
\sum_{\alpha=1}^{N}
\phi_{\alpha} f_{\alpha}(c_{\alpha})
+ \mu
\left[
c - \sum_{\alpha=1}^{N} \phi_{\alpha} c_{\alpha}
\right],
\end{equation}
where
$f_{\alpha}(c_{\alpha})$ is the bulk free energy density of phase or grain $\alpha$, $c_{\alpha}$ is the phase concentration of phase or grain $\alpha$, $c$ is the overall concentration in a given point within the system and $\mu$ is the Lagrange multiplier acting as a chemical potential enforcing mass conservation.

The evolution equation for the local concentration is given by the diffusion equation:

\begin{equation}
\dot{c} = \nabla \cdot \left[ \sum_{\alpha} \phi_{\alpha} \mathbf{D}^{\alpha} \nabla c_{\alpha}  + \sum_{\alpha, \beta} \mathbf{J}_{\alpha\beta}\right],
\end{equation}
where, $\mathbf{D}^{\alpha}$ denotes the diffusion coefficients matrix of phase $\alpha$ and $\mathbf{J}_{\alpha\beta}$ is the anti-trapping flux introduced to suppress numerical solute trapping associated with the diffuse-interface representation \cite{steinbach2009phase}. Both the phase-field evolution equation and the chemical diffusion equation are discretized on a regular grid using finite-difference scheme, with time integration performed via an explicit forward Euler method. 

The mechanical free-energy density, $f_{mech}$, quantifies the strain energy associated with deformation arising during the phase transformation. The material response is described using a St.~Venant–Kirchhoff hyperelastic formulation together with the Green–Lagrange finite-strain measure \cite{shchyglo2019phase,levitas2013phase,miehe2002strain}. The corresponding mechanical free-energy density is given by:

\begin{equation}
f_{\mathrm{mech}} =
\sum_{\alpha}^{el}
\phi_{\alpha}
\frac{1}{2}
\boldsymbol{E}_{\alpha}^{el}
:
\mathbb{C}_{\alpha}
:
\boldsymbol{E}_{\alpha}^{el},
\end{equation}
where $\boldsymbol{E}_{\alpha}^{el}$ is the elastic strain tensor of phase or grain $\alpha$ and $\mathbb{C}_{\alpha}$ is the corresponding elastic stiffness tensor.
 The Green-Lagrange finite strain tensor $\boldsymbol{E}_{\alpha}^{el}$ is given by:
 
 \begin{equation}
     \mathbf{E}_{\alpha}^{el} = \frac{1}{2} ((\mathbf{F^{el}})^{T}\mathbf{F^{el}} - \mathbf{I}),
 \end{equation}
 where $\mathbf{F^{el}}$ is the elastic deformation gradient tensor and $\mathbf{I}$ is the identity tensor. The elastic deformation gradient is described by the multiplicative decomposition of the total deformation gradient:
 
 \begin{equation}
     \mathbf{F} = \mathbf{F}^{el}\mathbf{F}^{pl}\mathbf{F}^{tr},
 \end{equation}
 where $\mathbf{F}^{pl}$ is the plastic deformation gradient tensor and $\mathbf{F}^{tr}$ is the transformation deformation gradient tensor. Thus, the elastic deformation gradient tensor reads:
 
 \begin{equation}
    \mathbf{F}^{el} = \mathbf{F}(\mathbf{F}^{pl}\mathbf{F}^{tr})^{-1}.
 \end{equation}
%
To study the flow behaviour of the material, a phenomenological Crystal Plasticity (CP) model was used. The evolution of the plastic deformation gradient $\mathbf{F}^{pl}$ is described in terms of the plastic velocity gradient $\mathbf{L}^{p}$ according to: 



\begin{equation}
\overset{.}{\textbf{F}}_p = \textbf{L}_p \textbf{F}_p,
\end{equation}
where $\dot{\textbf{F}}_p$ is the plastic deformation gradient rate. Accounting for the crystal symmetry, the plastic velocity gradient $\mathbf{L}_{p}$ is obtained by superposing the contributions from all active slip systems:

\begin{equation}
\textbf{L}_p = \sum_{s=1}^{N} \overset{.}{\gamma}^{s} \textbf{m}^{s} \otimes \textbf{n}^{s},
\end{equation}
where, $\textbf{m}^{s}$ and $\textbf{n}^{s}$ are unit vectors describing the slip direction and the normal direction to the slip plane of the slip system $s$. $\overset{.}{\gamma}$ is the shear rate of the slip system $s$ and N is the number of slip systems \cite{roters2010overview}.

Critical resolved shear stress, \textbf{$\tau_{c}^{s}$}, is used as a state variable for each slip system in phenomenological constitutive models. Shear rate is formulated as a function of resolved shear stress and critical resolved shear stress \cite{Peirce1983}. This model can also used to used study creep in Ni-based super alloys \cite{Ali2020}:
\begin{equation}
    \overset{.}{\gamma}^{s} = \overset{.}{\gamma}_{0} \left| \frac{{\tau^{s}}}{ \tau_{c}^{s}}\right| ^n sgn(\tau^{s}),
\end{equation} 
where $\dot{\gamma^{0}}$ is the reference shear rate and $n$ is the rate sensitivity exponent. The hardening law governs the evolution of the critical resolved shear stress as a  function of plastic strain rate \cite{kocks1975thermodynamics}:


\begin{equation}
    \overset{.}{\tau}^{\alpha}_{c} = \sum_{\beta = 1}^{N} h_{\alpha\beta}|\overset{.}{\gamma}^{\beta}|,
\end{equation}
\begin{equation}
    h_{\alpha\beta} = q_{\alpha\beta} \left[ h_{0} \left(1-\frac{\tau_{c}^{\beta}}{\tau_{s}} \right)^{a} \right],
\end{equation}
where, $\tau_{s}$ is the saturation stress, i.e., the upper limit of the attainable flow stress, $h_{0}$ denotes the hardening modulus governing the initial hardening rate, and $q_{\alpha\beta}$ is the latent hardening ratio controlling the interaction between slip systems $\alpha$ and $\beta$, the parameter $a$ is the hardening exponent that controls the rate at which saturation is approached. Note, that the phenomenological crystal plasticity formulation adopted here is not temperature dependent. The corresponding mechanical equilibrium problem for the coupled elasto-plastic response is solved using the FFT-based spectral algorithm proposed in \cite{shchyglo2024efficient}.

To realistically represent heat extraction during heat treatment, the thermal boundary condition is formulated using Newton’s law of cooling \cite{molnar1969newton}, augmented by a latent-heat source term to account for the latent heat released during the phase transformation:

\begin{equation}
\dot{T}(t) =
- r \left(T(t) - T_s \right)
+ \frac{Q}{\rho C_p} \dot{f},
\label{eq:CoolingLaw}
\end{equation}
where $T(t)$ denotes the sample temperature at time $t$, $T_s$ is the temperature of the cooling medium, $\rho$ is the mass density of the material, $C_p$ is the specific heat capacity of the material and $Q$ is the latent heat associated with the bainitic transformation. The term $\dot{f}$ represents the transformation rat averaged over the sample volume, i.e., the time derivative of the transformed phase fraction. The heat extraction coefficient $r$ reflects the cooling intensity imposed by the quenching medium (e.g., air, oil, or water). During solid-state transformations such as bainitic reactions, the release of latent heat can substantially influence the transformation kinetics and the resulting microstructure. As transformation proceeds, latent heat is released, which may locally slow down the cooling process or even cause recalescence. The modified thermal history feeds back into the thermodynamic driving force, thereby affecting growth rate of the newly formed phases. In practice, latent-heat release may retard the transformation and promote a more gradual microstructural evolution. Therefore, a direct coupling between the phase-field kinetics and the temperature evolution is essential to capture the transient thermal response of the transforming system.

The proposed phase-field framework accounts for the key physical mechanisms active during bainitic transformations, namely heat extraction including latent-heat effects, chemical diffusion, capillarity and mechanical relaxation. All of these contributions enter the transformation driving force.

\section{Simulation of bainitic microstructures}

\begin{figure}[H]
    \centering
    \includegraphics[width=1\textwidth]{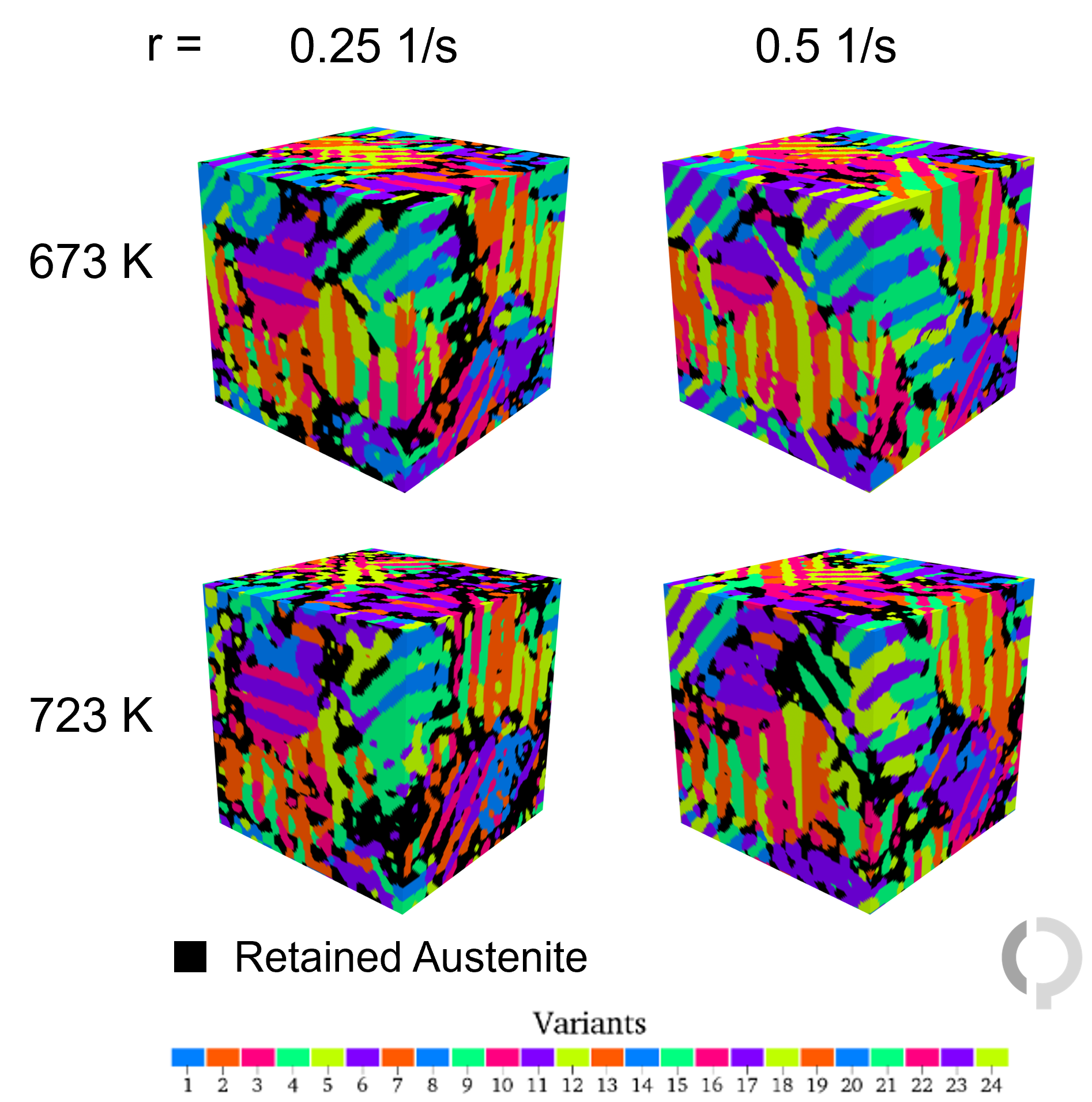}
    \caption{\label{fig:micro01}Simulated phase-field microstructures using different processing conditions }
\end{figure}

Three-dimensional phase-field simulations of bainite formation were carried out in a cubic domain of $\SI{128}{\micro\meter}$ size, discretized on a regular grid with a spacing, $\Delta x$, of  $\SI{0.1}{\micro\meter}$ and periodic boundary conditions in all directions. The diffuse-interface thickness was set to $5\Delta x$ and the interfacial properties were defined by an interface mobility of $\mu = 1\times10^{-13}\,\si{\meter^{4}\per\joule\second}$ and an interface energy of $\sigma_0=\SI{0.24}{\joule\per\meter\squared}$ (Table~\ref{tab:simulation_parameters_pf}). The simulations were initialized at 900 K, at which austenite is the only stable phase. Bainitic ferrite nuclei were introduced randomly into austenite containing $0.2$ wt.\% carbon, using a seed density of $7.5\times10^{17}\,\si{\meter^{-3}}$. Bainite formation was then simulated under two cooling bath temperatures, 673 K and 723 K, with temperature evolution governed by Newton’s law of cooling (Eq.~\ref{eq:CoolingLaw}) using two different heat-extraction coefficients $r = 0.25 $ and $0.5 s^{-1}$. The resulting microstructures can be seen in Figure~\ref{fig:micro01}.

\begin{table}[H]
\centering

\begin{tabular}{ll}
\hline
\textbf{Parameter} & \textbf{Symbol} \\
\hline

Domain size (x direction) & $L_{x} = \SI{128}{\micro\meter}$ \\
Domain size (y direction) & $L_{y} = \SI{128}{\micro\meter}$ \\
Domain size (z direction) & $L_{z} = \SI{128}{\micro\meter}$ \\
Grid spacing & $\Delta x = \SI{0.1}{\micro\meter}$ \\
Interface width & $\eta = 5  \Delta x$ \\
Interface mobility & $\mu = 1 \times 10^{-13}\,\si{\meter^{4}\per\joule\second}$ \\
Interface energy & $\sigma_{0} = \SI{0.24}{\joule\per\meter\squared}$ \\

\hline
\end{tabular}
\caption{Numerical parameters used in the simulations for this study}
\label{tab:simulation_parameters_pf}
\end{table}


\section{ Phenomenological yield function}

The onset of plastic deformation is reached when the applied stress state satisfies the yield condition of the material. Under uniaxial loading, the yield point can be identified directly from the stress-strain response. However, under multiaxial loading conditions, the transition from elastic to plastic behaviour cannot be described by a single stress component. Instead, plastic yielding is characterized by a function relating the principal stresses, commonly referred to as the yield function. The corresponding mathematical formulation is known as a yield criterion and defines the boundary between the elastic and plastic domains in stress space. This boundary is referred to as the yield surface, which also serves as a plastic potential for determining the direction of plastic strain evolution \cite{Banabic2010}.

 Barlat \cite{barlat91} provides one of the yield functions that account for six stress state components and is referred to Yld91.  The yield surface obtained from experiment and polycrystal model (Yld91) are found to be in reasonable agreement. The Yld91 yield function is given by \cite{barlat91,Zhang2016}:

\begin{equation}
\Phi = (3I_{2})^{n/2} \left\{ \left|2\,\cos\left( \frac{2\theta + \pi}{6}\right)\right|^{n} +\left|2\,\cos\left( \frac{2\theta + 3\pi}{6}\right)\right|^{n} + \left|2\,\cos\left( \frac{2\theta + 5\pi}{6}\right)\right|^{n}  \right\} = 2\sigma_{y}^{n},
\end{equation}
where $\theta = \arccos (I_{3}/I_{2}^{3/2})$. Equivalent stress is defined by $\phi = (\Phi/2)^{1/n}$, with $n= 6, 8$ for BCC and FCC materials, respectively. $\sigma_{y}$ is the yield stress along rolling direction.

\begin{eqnarray}
    I_{2} &=& \frac{(fF)^{2} + (gG)^{2} + (hH)^{2}}{3} + \frac{(aA - cC)^{2} + (cC - bB)^{2} + (bB - aA)^{2}}{54},\nonumber \\
    I_{3} &=& \frac{(aA - cC)(cC - bB)(bB - aA)}{54} + (fF)(gG)(hH) \nonumber \\
          &-& \frac{(cC - bB)(fF)^{2} + (aA - cC)(gG)^{2} + (bB - aA)(hH)^{2}}{6},
\end{eqnarray}
where $I_{2}$ and $I_{3}$ are second and third invariants of stress tensor. $A = \sigma_{22} - \sigma_{33}$, $B = \sigma_{33} - \sigma_{11}$, $C = \sigma_{11} - \sigma_{22}$, $F = \sigma_{23}$, $G = \sigma_{31}$, $H = \sigma_{12}$. $a,b,c,d,e,f$ are six anisotropic material parameters for Yld91 yield function. These material parameters can be obtained using inverse identification methods. One of the identification methods is the least square method, discussed in the following section.

\section{Results \& Discussion}

\begin{figure}[H]
    \centering
    \includegraphics[width=1\textwidth]{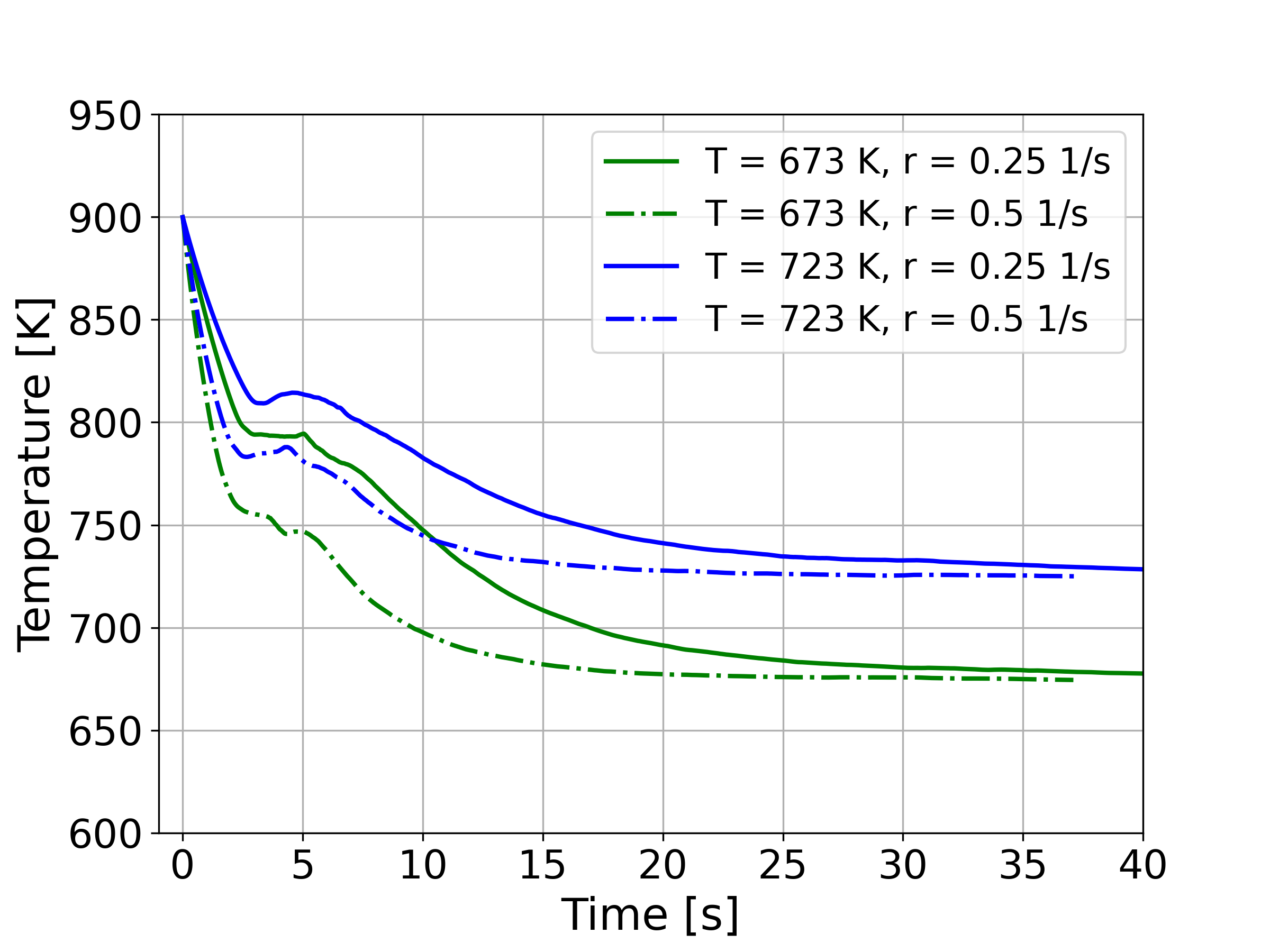}
    \caption{Cooling curves for different holding temperatures and heat extraction coefficients}
    \label{fig:cooling_curves}
\end{figure}

Figure~\ref{fig:cooling_curves} compares the thermal histories obtained for two cooling-bath temperatures, 673~K and 723~K, and two heat-extraction coefficients, $r=0.25~\si{\per\second}$ and $r=0.5~\si{\per\second}$. The larger magnitude, ($r=0.5\ \mathrm{s^{-1}}$), corresponds to a stronger heat extraction or faster cooling condition. All conditions start from the same initial temperature, however, the subsequent cooling response depends on both the bath temperature and the magnitude of the heat-extraction coefficient. The curves associated with the 673~K bath temperature exhibit a larger temperature reduction and approach a lower final temperature than those cooled towards 723~K. This is expected because the lower bath temperature provides a larger temperature difference between the sample and the cooling environment, resulting in a greater thermal driving force for heat removal.

For a given bath temperature, increasing the magnitude of the heat-extraction coefficient from $r=0.25~\si{\per\second}$ to $r=0.5~\si{\per\second}$ leads to a steeper initial decrease in temperature. Thus, the dash-dotted curves in Figure \ref{fig:cooling_curves} show more rapid cooling than the corresponding solid curves and reach their near-stationary temperature earlier. In contrast, the lower heat-extraction coefficient produces a more gradual cooling trajectory and prolongs the transient cooling period. The temporary temperature raise or plateau around $5\,s$ cooling time result from the latent-heat release during transformation, which counteracts the heat removal by the cooling medium. The results show that the bath temperature controls the final temperature, whereas the heat-extraction coefficient primarily controls the rate at which this temperature is approached.

\begin{figure}[H]
    \centering
    \includegraphics[width=1\textwidth]{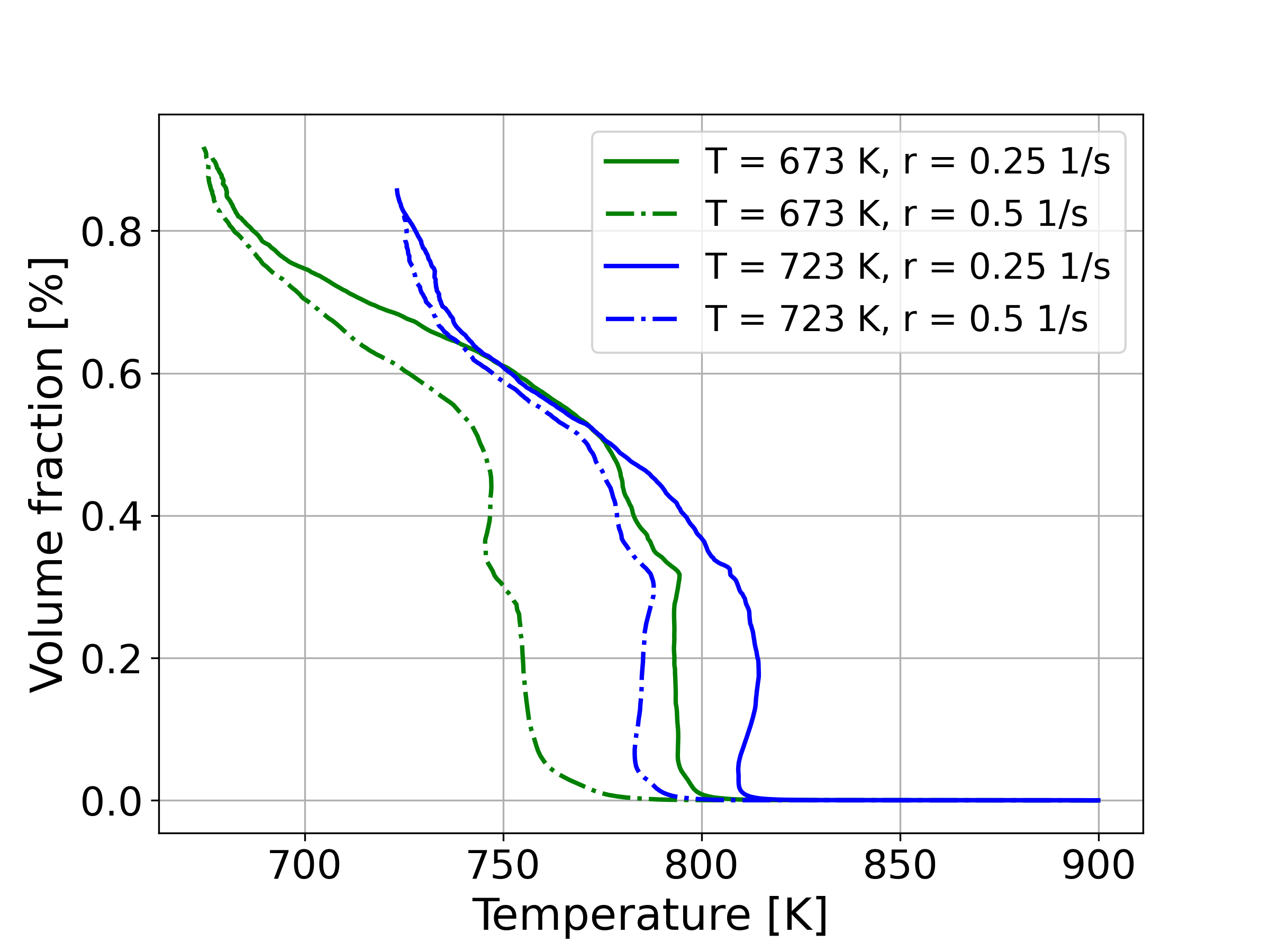}
    \caption{Effect of heat extraction coefficient and holding temperature on the bainitic ferrite volume fraction evolution during bainite formation}
    \label{fig:volume_frac}
\end{figure}

Figure~\ref{fig:volume_frac} presents the evolution of the bainitic ferrite (BF) volume fraction as a function of temperature for the four simulated processing conditions: holding temperatures of 673~K and 723~K combined with heat-extraction coefficients of $r=0.25~\si{\per\second}$ and $r=0.5~\si{\per\second}$. In all cases, the BF fraction increases as temperature decreases, indicating the progressive transformation of austenite into bainitic ferrite during cooling. The transformation begins at different temperatures for the two holding conditions and subsequently approaches a nearly constant final BF fraction as the reaction slows down.

The holding temperature has a pronounced effect on the transformation path and final transformed fraction. The curves obtained for the 723~K condition are generally shifted towards lower BF fractions compared with those for 673~K over the common temperature range. In particular, the condition with $T=723$~K and $r=0.25~\si{\per\second}$ reaches the lowest final BF volume fraction, whereas the 673~K conditions approach higher final BF volume fractions. This behaviour indicates that the thermal history associated with the lower holding temperature allows a greater fraction of bainitic ferrite to form within the simulated time interval.

The heat extraction coefficient has minor effect on the final BF volume fraction but is affecting the BF microstructure formation as will be discussed later in this work. Thus, at a given holding temperature, the lower heat-extraction coefficient, $r=0.25~\si{\per\second}$, results in a slightly lower BF volume fraction than the corresponding case with $r=0.5~\si{\per\second}$ but shows significant differences in the transformation kinetics. 

Overall, the results demonstrate that the bainitic transformation response is controlled by the combined effect of holding temperature and heat-extraction coefficient. The holding temperature primarily modifies the transformation trajectory and attainable BF fraction, whereas the heat-extraction coefficient determines how rapidly the thermal condition is reached and, consequently, the time available for bainite formation.

\subsection{Comparison of microstructures}

The simulated microstructure is subjected to a detailed morphological analysis to evaluate its features and characteristics. The results are then validated against experimental observations, such as those obtained from scanning electron microscopy (SEM). This comparison ensures the adequacy of the simulated microstructure by analyzing key parameters like bainite morphology and phase distribution. The insights gained through such a comparison help to refine the model for better alignment with real-world microstructural behavior. 

\subsubsection{Microstructure analysis from experiments}
\label{micro_expt}
\begin{figure}[H]
    \centering
    \includegraphics[width=1\textwidth]{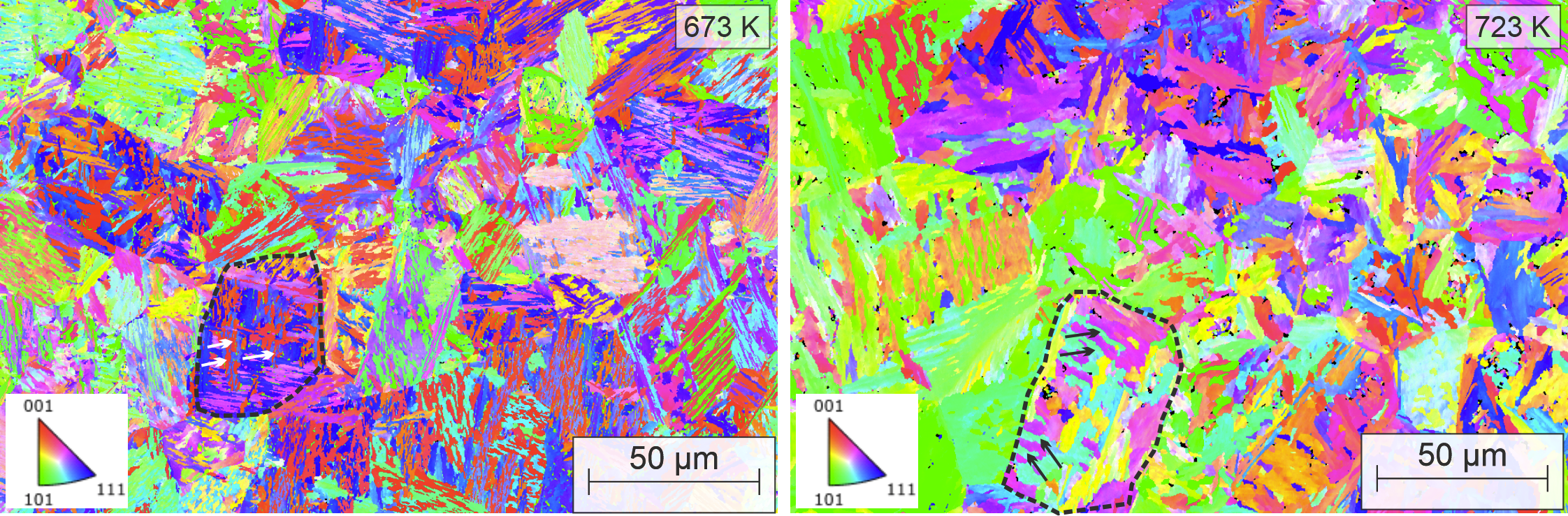}
    \caption{\label{fig:EBSD}Inverse pole figure maps of single prior austenite grain (PAG) from holding temperatures 673 K and 723 K, respectively.}
\end{figure}

For morphological feature analysis from experiments, the specimen surfaces were prepared by mechanical grinding down to 1200 grit SiC paper, followed by polishing with \SI{6}{\micro\meter} and \SI{1}{\micro\meter} diamond paste. The specimens were then etched with 3\% Nital. Scanning electron microscopy (SEM) was performed on a field-emission gun Zeiss Sigma microscope (Carl Zeiss Microscopy GmbH, Germany), operated with a \SI{30}{\micro\meter} aperture, an accelerating voltage of \SI{15}{\kilo\volt}, and a working distance of \SI{9}{\milli\meter} using the secondary electron (SE) detector. The resulting microstructures after isothermal holding at 673 K and 723 K are shown in Figure~\ref{fig:EBSD}. Both heat treatments resulted in bainitic microstructures. A clear difference in the scale of the bainitic ferrite plates is observed between the two conditions: the microstructure obtained at 723 K exhibits coarser plates as highlighted in dashed black box in Figure~\ref{fig:EBSD}, while the microstructure obtained at 673 K is finer, as also highlighted in the micrographs. Bainitic ferrite thickness of each variant is estimated for both heat treatment processes. And the corresponding statistics are presented in Table~\ref{tab:bf_width_validation}. The detailed extraction of features is presented in \cite{gulbay2023influence}.

\subsubsection{Statistical analysis of bainitic morphology from simulations}

\begin{figure}[H]
    \centering
    \includegraphics[width=1\textwidth]{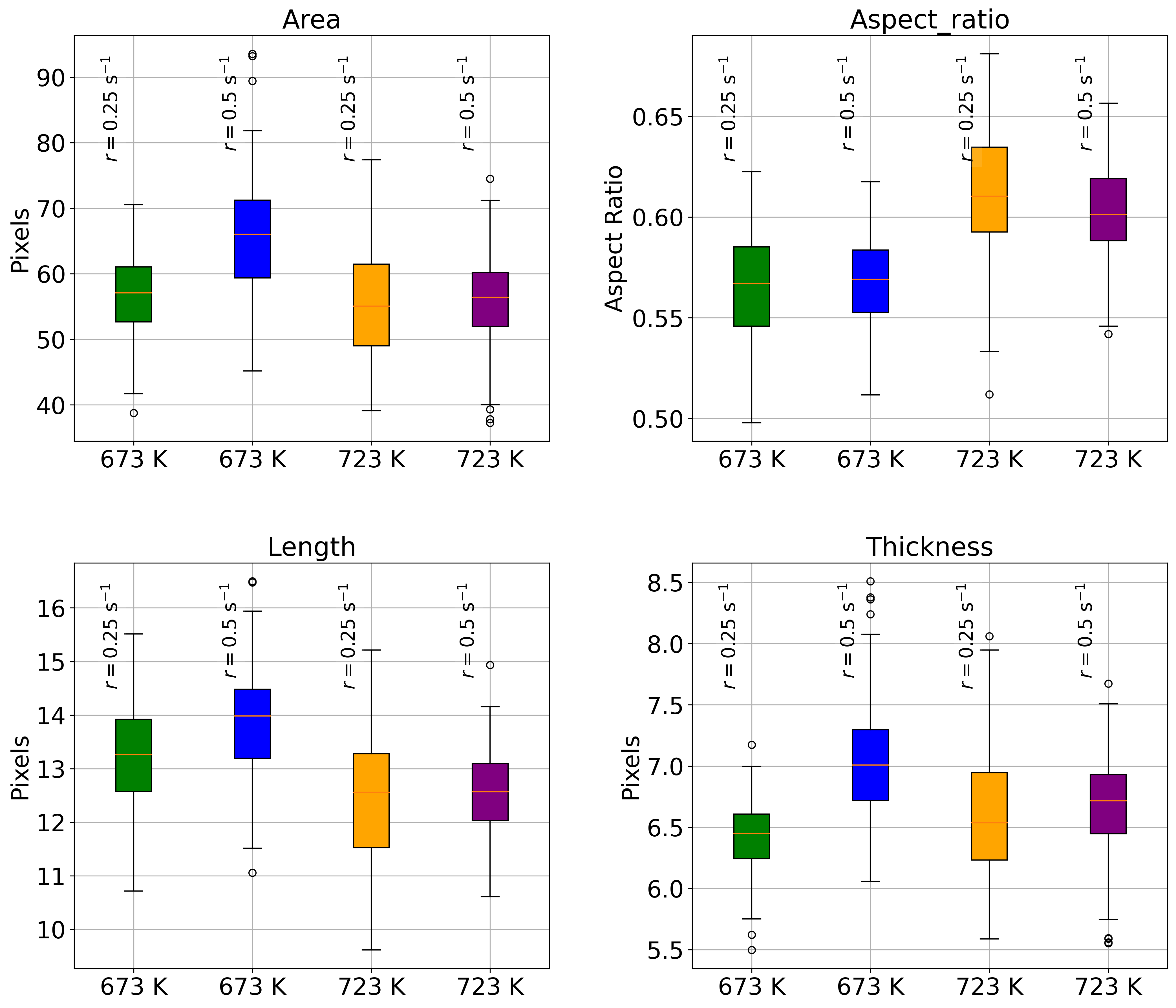}
    \caption{\label{fig:morph01}Statistical distributions of simulated bainitic ferrite morphology for the four thermal conditions: (a) area, (b) aspect ratio, (c) length, and (d) thickness. The results compare holding temperatures of 673~K and 723~K combined with heat-extraction coefficients of $r=0.25~\mathrm{s^{-1}}$ and $r=0.5~\mathrm{s^{-1}}$.}
\end{figure}

Figure~\ref{fig:morph01} compares the distributions of bainitic ferrite area, aspect ratio, length, and thickness for the four simulated thermal conditions. The results demonstrate that both the holding temperature and the heat-extraction coefficient influence the characteristic size and shape of the transformed bainitic regions. The response is not uniform for all morphological descriptors, indicating that the simulated transformation is governed by a competition between nucleation, growth and mechanical accommodation rather than by a single thermal parameter.

The area distribution shown in Fig.~\ref{fig:morph01}(a) indicates that the condition at $673~\mathrm{K}$ with $r=0.5~\mathrm{s^{-1}}$ produces the largest bainitic ferrite features among the investigated cases. Increasing the heat-extraction coefficient at $673~\mathrm{K}$ therefore promotes the formation of larger transformed regions in the present simulations. In contrast, the two conditions at $723~\mathrm{K}$ exhibit comparatively smaller bainitic areas, with only a limited difference between $r=0.25~\mathrm{s^{-1}}$ and $r=0.5~\mathrm{s^{-1}}$. This suggests that, at the higher holding temperature, the morphology is less sensitive to the applied heat-extraction condition and is mostly controlled by the reduced chemical driving force.

The aspect-ratio distributions in Fig.~\ref{fig:morph01}(b) show a clear effect of holding temperature on the shape of the bainitic features. The conditions simulated at $723~\mathrm{K}$ exhibit higher aspect-ratio values than those at $673~\mathrm{K}$. The figure indicates that bainitic regions formed at $723~\mathrm{K}$ are comparatively coarser and more equiaxed, whereas the lower aspect-ratio values at $673~\mathrm{K}$ correspond to more elongated features. The lower holding temperature provides a higher thermodynamic driving force for transformation, which favours elongation of bainitic ferrite. At $723~\mathrm{K}$, the reduced driving force and enhanced diffusion-related relaxation reduce the tendency for strongly anisotropic growth. In particular, the condition at $723~\mathrm{K}$ with the lower heat-extraction coefficient, $r=0.25~\mathrm{s^{-1}}$, provides a longer effective thermal exposure and therefore greater time for lateral growth and coarsening of the bainitic regions. Consequently, this condition promotes the development of comparatively larger and more compact features than the corresponding higher heat-extraction condition. The influence of the heat-extraction coefficient on aspect ratio is nevertheless smaller than the influence of holding temperature, particularly for the $723~\mathrm{K}$ conditions.

The length distributions in Fig.~\ref{fig:morph01}(c) are consistent with the area results. The bainitic features formed at $673~\mathrm{K}$ with $r=0.5~\mathrm{s^{-1}}$ exhibit the characteristic length, followed by the $673~\mathrm{K}$ condition with $r=0.25~\mathrm{s^{-1}}$. In contrast, both $723~\mathrm{K}$ conditions show shorter bainitic features. These results indicate that the combination of lower holding temperature and higher heat extraction favours elongated growth of bainitic ferrite in the simulated domain.

The thickness distributions in Fig.~\ref{fig:morph01}(d) show only limited variation among the simulated conditions, with most bainitic features exhibiting thicknesses of approximately 6-7 pixels. From a physical perspective, the condition at $673~\mathrm{K}$ with $r=0.5~\mathrm{s^{-1}}$ would be expected to produce the smallest bainitic ferrite thickness because the lower holding temperature and stronger heat extraction should restrict lateral growth and favour a finer transformation product. However, this expected refinement trend is not clearly resolved in the morphology statistics. This limited sensitivity is attributed primarily to the spatial resolution of the present simulations and the image-based thickness evaluation, since the measured feature thickness is represented by only a small number of grid cells. As a result, small differences in lateral growth may lie within the discretization uncertainty and cannot be distinguished reliably from the boxplot distributions alone. Therefore, the thickness statistics in Fig.~\ref{fig:morph01}(d) should be interpreted cautiously and are further assessed through comparison with experimentally measured bainitic ferrite widths.

Overall, the morphological statistics indicate that the thermal pathway strongly affects the simulated bainitic transformation, particularly with respect to feature area, length, and aspect ratio. The most pronounced variation is observed between the two conditions at $673~\mathrm{K}$, where increasing the heat-extraction coefficient from $r=0.25~\mathrm{s^{-1}}$ to $r=0.5~\mathrm{s^{-1}}$ promotes larger and more elongated bainitic features in the present simulations. In contrast, the conditions at $723~\mathrm{K}$ produce comparatively more compact morphologies and exhibit a weaker sensitivity to the heat-extraction coefficient. Although the thickness trend cannot be resolved clearly from the pixel-based morphology statistics, the subsequent comparison with experimental bainitic ferrite widths confirms the expected temperature-dependent thickening behaviour. Since there are no experimental estimations of area and length distributions,it should be noted that the area and length distributions which reported here are extracted directly from phase-field simulation microstructures and could not be compared to experimental trends. Overall, the simulated morphology reflects the coupled influence of transformation driving force, interface mobility, carbon redistribution, and mechanical accommodation incorporated in the phase-field model. 

The bainitic ferrite thickness obtained from the phase-field simulations was compared with experimentally measured values, as summarized in Table~\ref{tab:bf_width_validation}. Both the experiments and simulations show an increase in bainitic ferrite thickness with increasing holding temperature from $673~\mathrm{K}$ to $723~\mathrm{K}$. This trend is consistent with the kinetics of bainitic transformation. At $673~\mathrm{K}$, the larger undercooling increases the thermodynamic driving force for bainite formation and promotes the formation of finer bainitic ferrite features. In contrast, at $723~\mathrm{K}$, the lower driving force together with the longer time available for carbon diffusion and mechanical accommodation allows the bainitic ferrite plates to grow and thicken more gradually, resulting in a larger characteristic thickness.

Quantitatively, the phase-field simulations estimates bainitic ferrite thickness of $0.341~\mu\mathrm{m}$ at $673~\mathrm{K}$ and $0.382~\mu\mathrm{m}$ at $723~\mathrm{K}$. These values exceed the experimentally measured mean thickness of $0.15~\mu\mathrm{m}$ and $0.26~\mu\mathrm{m}$, respectively. The deviation is more pronounced at $673~\mathrm{K}$, where the predicted thickness is above the experimental range of $0.10\text{--}0.20~\mu\mathrm{m}$. At $723~\mathrm{K}$, the simulated thickness is closer to the experimental range of $0.18\text{--}0.34~\mu\mathrm{m}$, although it remains slightly above the upper limit. Thus, the phase-field model reproduces the experimentally observed temperature-dependent trend in bainitic ferrite thickness, while overestimates the absolute ferrite thickness, particularly for the lower holding-temperature condition. This comparison indicates qualitative agreement between simulation and experiment. Further refinement of phase-field simulations like spatial resolution is required to improve qualitative agreement between simulated and experimental microstructure morphology.

\begin{table}[htbp]
\centering

\begin{tabular}{c|c|c}
\hline
Holding temperature & Experiment & Phase-field simulation \\
\hline
\(673~\mathrm{K}\) & \(0.15 \pm 0.05~\mu\mathrm{m}\) & \(0.341~\mu\mathrm{m}\) \\
\(723~\mathrm{K}\) & \(0.26 \pm 0.08~\mu\mathrm{m}\) & \(0.382~\mu\mathrm{m}\) \\
\hline
\end{tabular}
\caption{Comparison of bainitic ferrite (BF) thickness obtained from experiments \cite{gulbay2023influence} and phase-field simulations.}
\label{tab:bf_width_validation}
\end{table}

\subsection{Internal stresses after bainitic transformation}

\begin{figure}[H]
    \centering
    \includegraphics[width=1\textwidth]{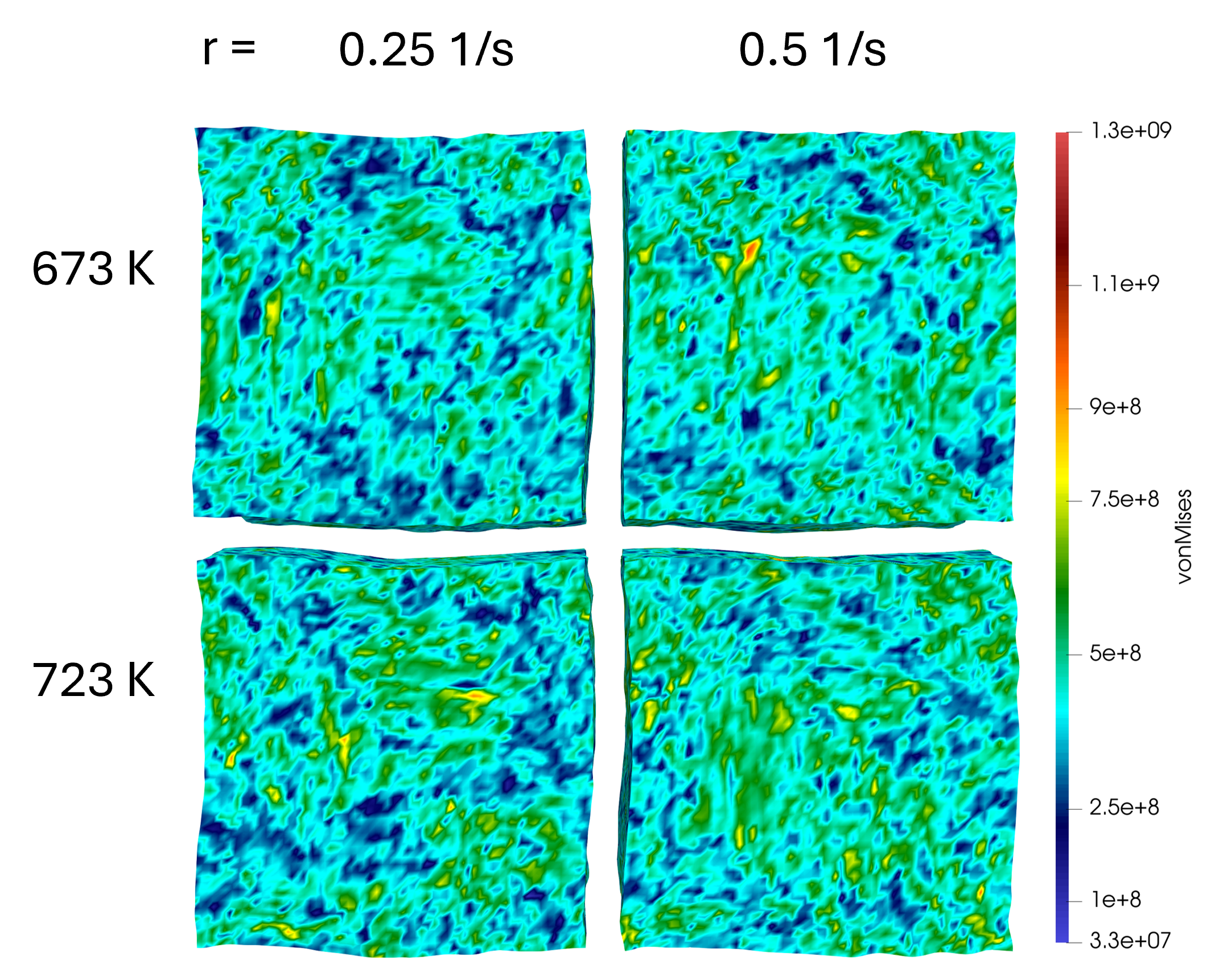}
    \caption{\label{fig:residual_stress_micro} Internal von Mises stresses after the phase transformation}
\end{figure}

The spatial stress maps in Fig.~\ref{fig:residual_stress_micro} demonstrate that the internal stress state is highly heterogeneous. For both holding temperatures, increasing the magnitude of the heat extraction coefficient from \(r=0.25~\mathrm{s^{-1}}\) to \(r=0.5~\mathrm{s^{-1}}\) accelerates the stress build-up and increases the maximum internal von Mises stress. The \(673~\mathrm{K}\), \(r=0.5~\mathrm{s^{-1}}\) condition exhibits the highest stresses, indicating that faster heat extraction promotes a more rapid transformation event. Under this condition, transformation-induced strain is introduced over a short time interval and cannot be fully accommodated by plastic relaxation, resulting in a pronounced accumulation of elastic strain energy. In contrast, the \(r=0.25~\mathrm{s^{-1}}\) condition produces lower maximum stress, indicating that the slower thermal evolution provides more time for mechanical accommodation during bainite growth.

The effect of holding temperature is also evident. At \(723~\mathrm{K}\), the stress increase is generally more gradual and the peak stress remains lower than that obtained at \(673~\mathrm{K}\) under the faster heat extraction condition. At the higher holding temperature, the reduced transformation driving force and the longer available time for local plastic accommodation limit the rate of stress accumulation. In contrast, the larger undercooling at \(673~\mathrm{K}\) increases the magnitude of the chemical driving force for transformation and promotes a more rapid generation of transformation-induced strain. Consequently, the balance between stress generation and plastic relaxation shifts towards stronger transient stress accumulation.

Rather than being uniformly distributed throughout the microstructure, the internal von Mises stress is concentrated in local regions. These local stress concentrations are associated with microstructural incompatibilities generated during bainite formation, including transformation strain mismatch between neighbouring bainitic regions and incomplete local plastic accommodation. Localised stress hot spots may strongly influence subsequent yielding, strain localisation and potential damage initiation during mechanical loading.

It should be noted that the local von Mises stresses shown in Fig.~\ref{fig:residual_stress_micro} represent internally constrained microstructural stresses rather than directly measurable macroscopic tensile stresses. Their high magnitudes should therefore be interpreted as indicators of transformation-induced mechanical incompatibility and local elastic constraint. Since damage, cracking and other stress-relaxation mechanisms are not explicitly included in the present model, the absolute local stress values should be interpreted with caution. Nevertheless, the relative differences between heat-treatment conditions provide valuable insight into the coupling between heat extraction, transformation kinetics, microstructural evolution, and internal-stress development.

The internal-stress analysis demonstrates that stronger heat extraction promotes more rapid bainitic transformation and higher transient internal stresses, whereas higher holding temperature facilitates stress relaxation and reduces the severity of stress accumulation. These results confirm that the internal stress state is controlled by the coupled interaction of thermal driving force, transformation kinetics, local microstructural constraint and plastic relaxation.


\subsection{Tensile tests from experiments}

Uniaxial tensile tests were carried out on B5$\times$25 cylindrical specimens with a gauge diameter of \SI{5}{\milli\meter} and a gauge length of \SI{25}{\milli\meter}. The tests were conducted at room temperature on a universal electromechanical testing machine (ZwickRoell Z100, ZwickRoell GmbH \& Co.\ KG, Germany) at a strain rate of \SI{0.001}{\per\second}. The experimental stress-strain curves can be seen in Figure~\ref{fig:comptt}. As the isothermal holding temperature was lowered from \SI{723}{\kelvin} to \SI{673}{\kelvin}, the yield strength increased from \SI{667}{\mega\pascal} to \SI{765}{\mega\pascal}. It is evident that the mechanical properties became superior with decreasing the holding temperature. The higher yield strength and hardness of Carbide Free Bainite (CFB) - 673 K can be attributed to the increased fraction of the stronger bainitic ferrite (BF) and to the microstructural refinement resulting from the lower holding temperature. Yield strength for respective heat treatments can be seen in Table~\ref{tab:yield_strength}.

\begin{table}[htbp]
\centering
\begin{tabular}{c|c}
\hline
\textbf{Heat treatment state} & \textbf{Yield strength (MPa)} \\
\hline
\SI{673}{\kelvin} & $765 \pm 5$ \\
\SI{723}{\kelvin} & $667 \pm 5$ \\
\hline
\end{tabular}
\caption{Experimental yield strength of bainitic steels subjected to different isothermal heat treatment temperatures.}
\label{tab:yield_strength}
\end{table}

\subsection{Tensile response after different heat-treatment conditions}

Following the phase-transformation simulations, the transformed microstructures contain non-uniform transformation-induced internal stresses. 
The tensile simulations were then continued under monotonic loading in the \(x\)-direction using the same crystal-plasticity constitutive parameters for all heat-treatment conditions. The resulting stress-strain curves were subsequently compared with the experimental tensile data, as discussed below.

\begin{figure}[H]
    \centering
    \includegraphics[width=1\textwidth]{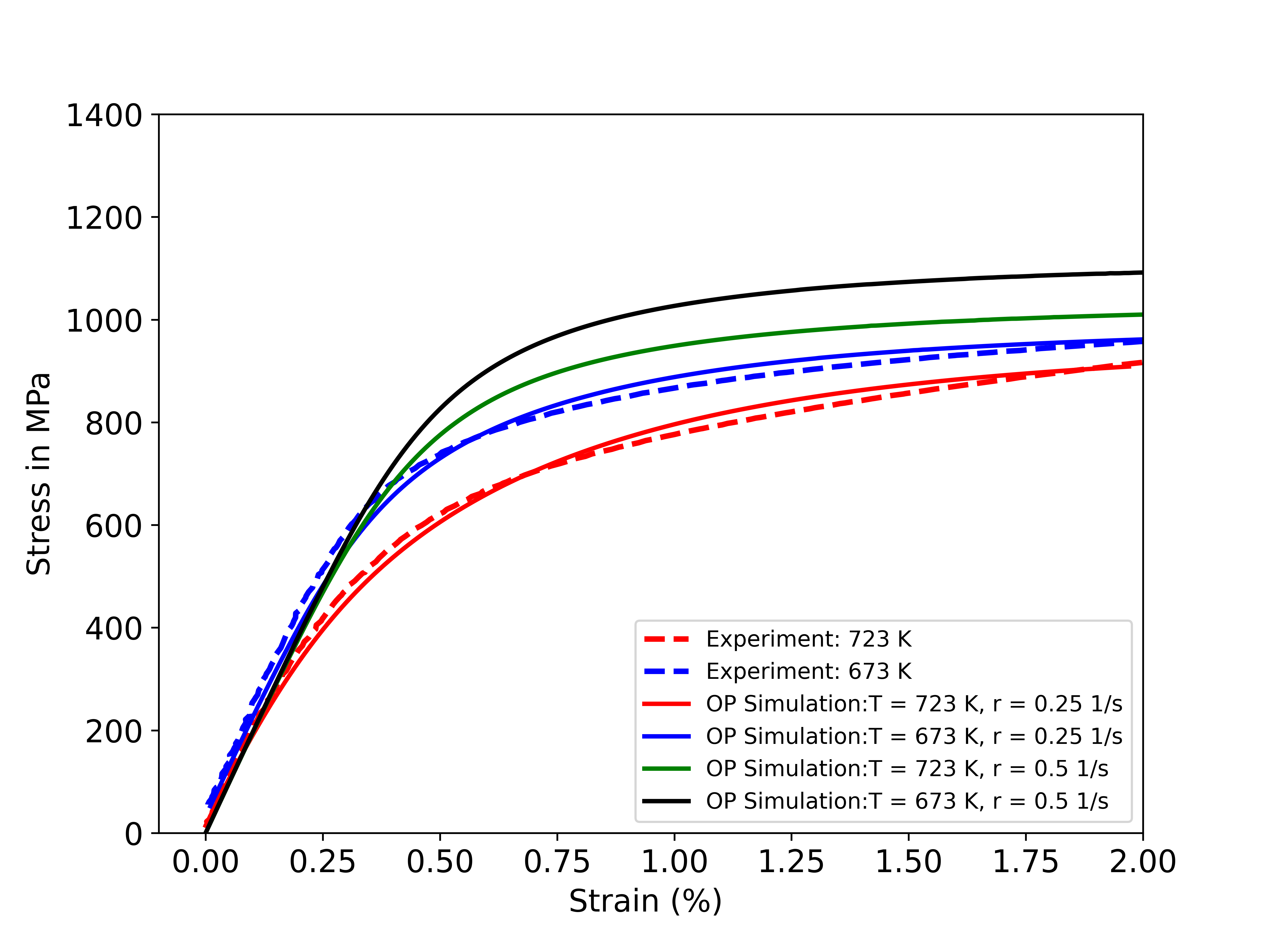}
    \caption{\label{fig:comptt}Comparison of tensile test results between PF simulation and experiments for two holding temperatures}
\end{figure}

The simulated and experimental tensile responses after bainitic heat treatment are shown in Fig.~\ref{fig:comptt}. The experimental curves show that the specimen transformed at \(673~\mathrm{K}\), exhibits a higher flow stress than the specimen transformed at \(723~\mathrm{K}\).~ 
The phase-field-based tensile simulations reproduce the same qualitative temperature dependence. For both heat extraction coefficients, the simulated \(673~\mathrm{K}\) curves show higher stresses than the corresponding \(723~\mathrm{K}\) curves. This indicates that the simulated bainitic microstructures capture the experimentally observed strengthening effect of lower holding temperature. Among the simulated cases, the curves obtained with \(r=0.25~\mathrm{s^{-1}}\) show the closest agreement with the experimental tensile curves. In particular, the \(673~\mathrm{K}\), \(r=0.25~\mathrm{s^{-1}}\) simulation follows the experimental \(673~\mathrm{K}\) response closely, while the \(723~\mathrm{K}\), \(r=0.25~\mathrm{s^{-1}}\) simulation also agrees well with the experimental \(723~\mathrm{K}\) curve.

In comparison, the higher heat extraction coefficient, \(r=0.5~\mathrm{s^{-1}}\), represents a faster cooling condition. The stronger heat extraction increases the effective undercooling rate and accelerates the bainitic transformation kinetics. As a result, the bainitic ferrite width becomes finer. Such refinement increases the number of barriers and therefore enhances the tensile strength. In addition, faster transformation can generate larger transformation-induced internal stresses because the transformation strain is introduced over a shorter time interval. If plastic accommodation cannot fully relax these stresses during transformation, the remaining internal stress state further contributes to the increased apparent flow stress during tensile loading.

Thus, the \(r=0.5~\mathrm{s^{-1}}\) simulations should be interpreted as a faster heat extraction case that produces a stronger and finer bainitic microstructure than the experimentally closer \(r=0.25~\mathrm{s^{-1}}\) condition. The comparison demonstrates that lower transformation temperature and faster heat extraction both increase strength, but through coupled mechanisms: enhanced chemical driving force, accelerated transformation kinetics, bainitic ferrite refinement and transformation-induced internal stress development. The tensile results support the transformation-microstructure-property relationship established in the previous sections, while confirming that \(r=0.25~\mathrm{s^{-1}}\) provides the best agreement with the experimental tensile response.

\begin{table}[]
\centering
\begin{tabular}{llll}
\hline
\textbf{Parameter} & \textbf{Symbol} & \textbf{Value} & \textbf{Unit} \\
\hline

Latent Hardening & $q_{t,u}$ & 1.4 & $-$ \\
Reference slip rate & $ \dot{\gamma_{0}}$ & 3e-4 & $s^{-1}$\\
Strain rate sensitivity & $n$ & 7 & $-$ \\
Critical resolves shear stress & $ \tau_{c} $ & 4.5e8 & $MPa$ \\
Saturated resolves shear stress & $ \tau_{s} $ & 6.0e8 & $MPa$ \\
Initial hardening & $h_{0}$ & 1e9 & $GPa$\\
Hardening Index & m & 2.0 & $-$ \\

\hline
\end{tabular}

\caption{Simulation parameters used for phenomenological crystal plasticity model}
\label{tab:simulation_parameters_cp}
\end{table}

\subsection{Yield surface estimation using Barlat91 phenomenological yield function.}

\begin{figure}[H]
    \centering
    \includegraphics[width=1\textwidth]{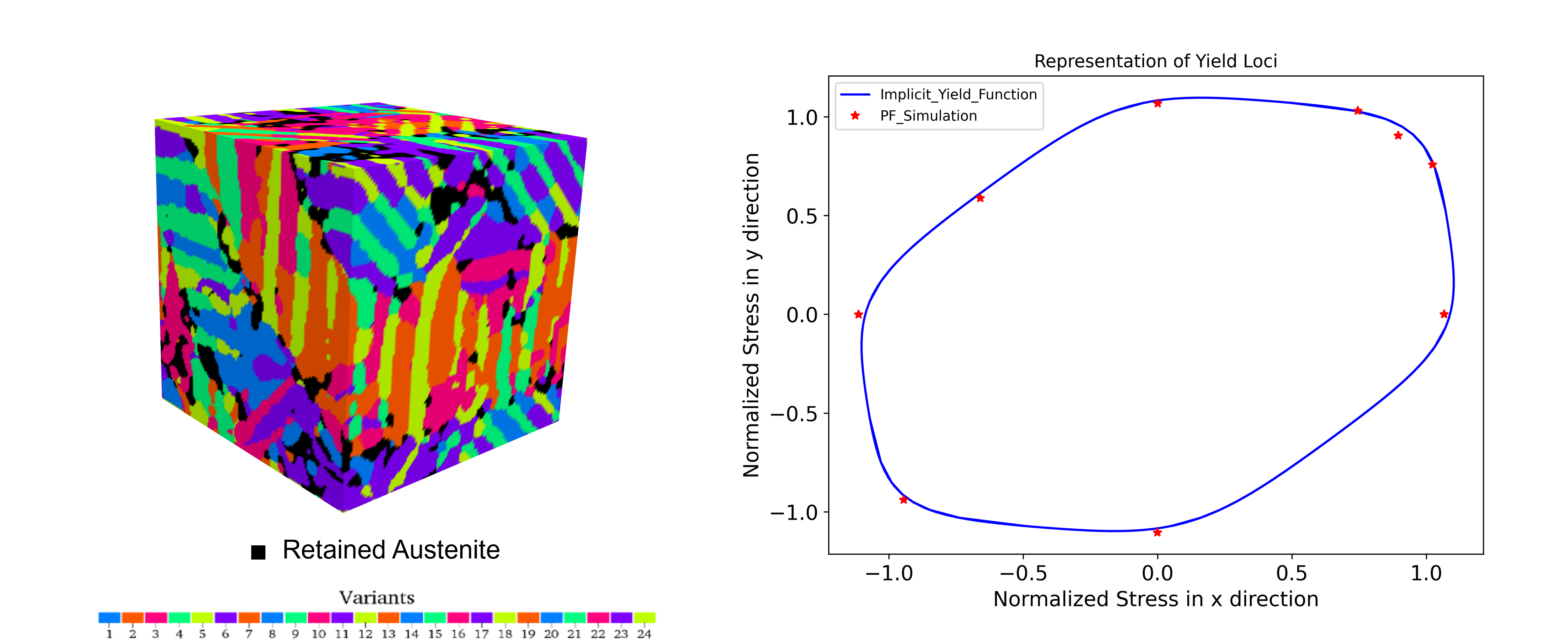}
    \caption{\label{fig:single_yield}Comparison of implicit yield surfaces for the microstructure holding at 673K with heat extraction coefficient of 0.25}
\end{figure}

Figure~\ref{fig:single_yield} compares the yield locus obtained from the fitted Barlat91 phenomenological yield function with the yield points extracted directly from the phase-field/crystal-plasticity simulations for the same bainitic microstructure. The uniaxial tensile response in the \(x\)-direction was previously validated against the available experimental tensile data and was used as the primary reference for calibrating the mechanical response. Subsequently, the phase-field-generated microstructure was subjected numerically to additional loading directions, including tension-compression and shear-dominated loading paths, while retaining the same crystal-plasticity parameter set. The resulting discrete yield points were then used to construct the macroscopic Barlat91 yield surface. The phenomenological Barlat91 locus provides an adequate representation of the anisotropic yield response estimated by the microstructure-resolved simulations.

\begin{figure}[H]
    \centering
    \includegraphics[width=1\textwidth]{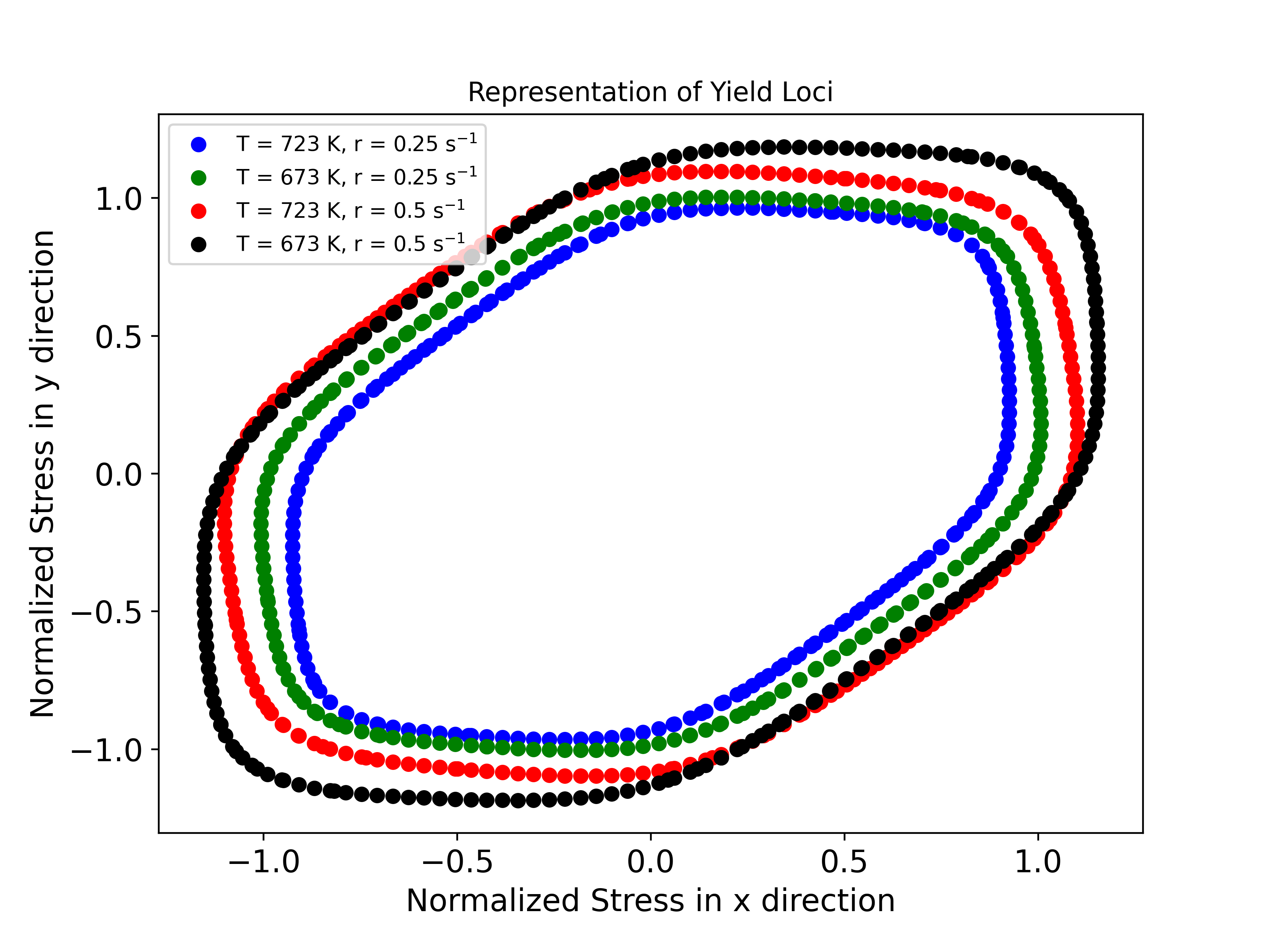}
    \caption{\label{fig:compyield}Comparison of implicit yield surfaces from the four heat treatment processes}
\end{figure}

The yield surfaces obtained for four different microstructures using the Barlat91 phenomenological yield function are shown in Fig.~\ref{fig:compyield}. The yield loci are clearly non-circular, indicating an anisotropic plastic response in the simulated bainitic microstructures. The deviation from a circular von Mises-type yield surface demonstrates that the plastic response is direction-dependent and cannot be fully represented by an isotropic plasticity model. Among the four simulated heat-treatment conditions, the \(673~\mathrm{K}\), \(r=0.5~\mathrm{s^{-1}}\) case exhibits the largest yield surface, followed by the \(723~\mathrm{K}\), \(r=0.5~\mathrm{s^{-1}}\) condition. 


The expansion of the yield surface with increasing heat extraction coefficient can be directly linked to the transformation kinetics and the resulting bainitic morphology. Faster heat extraction increases the effective cooling intensity and accelerates the development of thermodynamic driving force for bainite formation. Consequently, the transformation proceeds more rapidly, producing finer bainitic ferrite features. These microstructural features act as barriers to plastic deformation and increase the resistance to yielding. In addition, rapid transformation can generate higher transformation-induced internal stresses, since the transformation strain is introduced over a shorter time interval and may not be fully accommodated by plastic relaxation. The combined effect of bainitic ferrite refinement, increased microstructural constraint and retained internal stress leads to the outward expansion of the yield surface for the \(r=0.5~\mathrm{s^{-1}}\) cases.

The influence of the holding temperature is also evident from the relative size of the yield loci. For a given heat extraction coefficient, the \(673~\mathrm{K}\) condition generally shows a larger yield surface than the corresponding \(723~\mathrm{K}\) condition. This trend is consistent with the previously discussed morphology and tensile-test results. At \(673~\mathrm{K}\), the larger undercooling increases the magnitude of the chemical driving force for bainitic transformation and promotes finer and more elongated bainitic features. These features enhance the resistance to plastic flow. In contrast, transformation at \(723~\mathrm{K}\) provides more time for plastic accommodation and bainitic ferrite thickening, resulting in a comparatively lower resistance to plastic deformation. Therefore, the yield surfaces support the process-microstructure-property relationship observed from the phase-field simulations and tensile tests.

The anisotropy observed in the yield loci originates from the heterogeneous bainitic microstructure, including the morphology, variant orientation, spatial distribution of bainitic ferrite plates and transformation-induced internal stress field. Since bainite forms with directional features and variant-dependent transformation strains, the resulting microstructure does not resist plastic flow equally in all loading directions. Loading parallel to the dominant bainitic features may activate different slip systems and generate different constraint conditions compared with loading perpendicular to the bainitic plates. Similarly, shear-dominated loading states interact with the internal-stress field and phase-boundary network differently from uniaxial tensile loading. Therefore, the shape of the yield surface reflects both crystallographic anisotropy captured by the crystal-plasticity formulation and morphological anisotropy generated during the phase transformation.

The reduced separation between the yield surfaces in the shear-dominated regions of the yield locus indicates that the influence of holding temperature on the macroscopic response becomes less pronounced under shear-like loading. Although the two conditions processed with $r=0.5~\mathrm{s^{-1}}$ exhibit different overall yield-surface sizes, their loci approach each other in selected shear-dominated stress states. This behaviour can be related to the collective contribution of crystallographic bainitic ferrite variants rather than to the response of individual bainitic plates. Under uniaxial loading, the yield stress is strongly affected by the alignment of bainitic ferrite laths, transformation strains, and local internal-stress fields relative to the loading axis. In contrast, shear-dominated loading activates slip systems over a broader range of crystallographic orientations and therefore reduces the influence of a single preferred growth direction or morphological alignment.

\begin{figure}[H]
    \centering
    \includegraphics[width=1\textwidth]{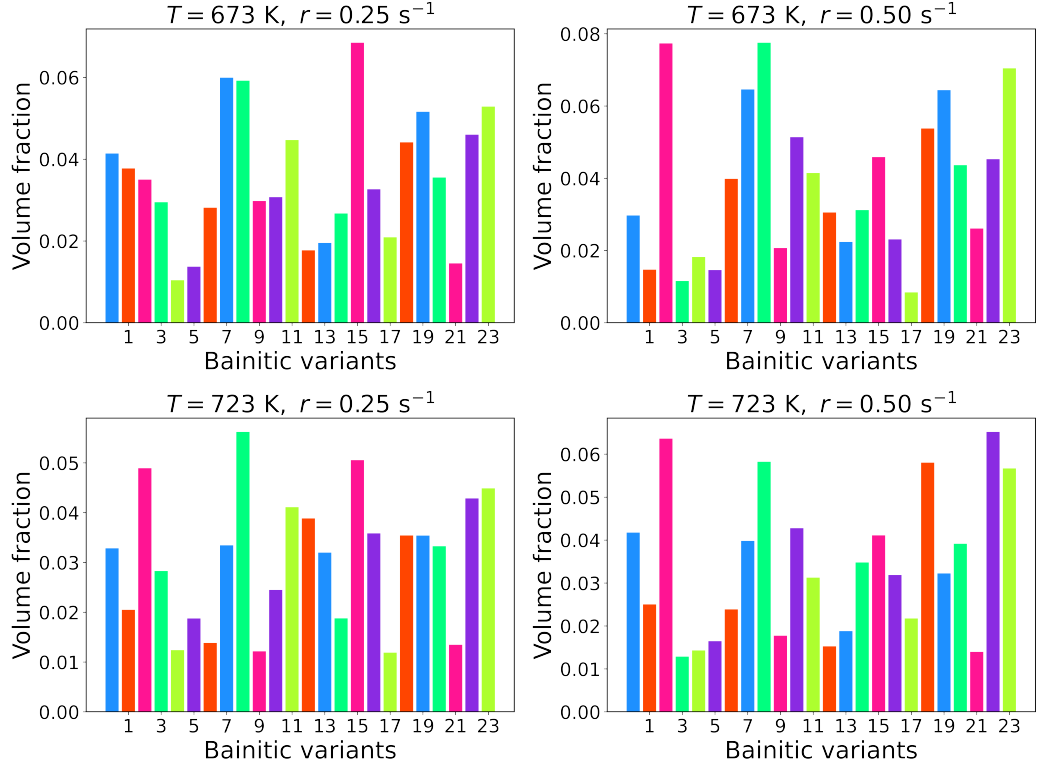}
    \caption{\label{fig:variants}Volume fractions of the 24 Kurdjumov-Sachs(KS) bainitic ferrite variants after transformation under different heat treatment conditions. Variant indices 0-23 correspond to the crystallographic variants defined according to the KS orientation relationship.}
\end{figure}

For the $r=0.5~\mathrm{s^{-1}}$ conditions, the variant-volume-fraction distributions provide a possible explanation for the similar shear response. Both the $673~\mathrm{K}$ and $723~\mathrm{K}$ microstructures contain comparable fractions of several Kurdjumov-Sachs related variants  shown in Figure~\ref{fig:variants}, particularly variants 2, 4, 9, 10, 13, 15, 20, and 23. Since these variants possess defined crystallographic orientations and associated transformation strains with respect to the parent austenite, their relative volume fractions influence the resolved shear stresses and the activation of slip systems during deformation. Comparable fractions of variants with similar shear orientations can therefore lead to a similar volume-fraction-weighted resistance to shear, even when the two microstructures differ in morphology, total strength level, or the relative fractions of other variants. The similar yield loci in the shear-dominated region should therefore be interpreted as the result of partial averaging of variant-level deformation behaviour and self-accommodation effects within the bainitic variant network.

In the present work, the tensile response in the $x$-direction was calibrated and validated using experimental tensile data, whereas the remaining loading directions were estimated using the same crystal-plasticity parameter set and the phase-field generated microstructures. This procedure provides a physically consistent estimate of the anisotropic yield response because all loading paths are evaluated using identical constitutive assumptions, phase distributions and variant orientations. Nevertheless, the response outside the experimentally validated tensile direction represents a model-based extrapolation. 

The Barlat91 yield criterion is suitable for representing the resulting anisotropic macroscopic response because it captures anisotropic yielding. In comparison with the isotropic von Mises criterion, Barlat91 accounts for the effect of processing-induced texture and morphology on yield-surface shape. It also provides greater flexibility than the quadratic Hill48 formulation for representing non-quadratic anisotropy. At the same time, it offers a computationally efficient continuum-scale representation of the yield behaviour obtained from the microstructure-resolved simulations. However, Barlat91 remains phenomenological and does not explicitly represent variant-level slip activity, local phase-boundary constraints, dislocation evolution, or redistribution of transformation-induced internal stresses. Despite these limitations, the fitted yield surfaces provide a useful macroscopic description of how heat treatment modifies anisotropic yielding through the combined effects of bainitic morphology, internal stresses, and the relative volume fractions of Kurdjumov-Sachs variants.

\section{Conclusions}

In this work, a phase-field-based framework was used to investigate the effect of bainitic heat-treatment conditions on transformation kinetics, microstructural morphology, internal-stress evolution, tensile response and anisotropic yielding behaviour. The main conclusions are summarized as follows:

\begin{itemize}

    \item The heat-treatment process strongly controls the austenite-to-bainite transformation kinetics. The cooling curves and phase-fraction evolution show that both holding temperature and heat extraction coefficient influence the transformation pathway. The lower holding temperature of \(673~\mathrm{K}\) promotes faster bainitic transformation due to the larger undercooling and the corresponding increase in the magnitude of the chemical driving force. In contrast, transformation at \(723~\mathrm{K}\) proceeds more gradually because the reduced driving force delays the transformation and allows more time for diffusion-assisted relaxation and plastic accommodation.

    \item The heat extraction coefficient modifies the transformation kinetics and final microstructural state. The condition \(r=0.25~\mathrm{s^{-1}}\) is closest to the experimental cooling route and provides the best agreement with the experimentally validated tensile response. The higher heat extraction coefficient, \(r=0.5~\mathrm{s^{-1}}\), represents a faster cooling condition. The stronger heat extraction accelerates the transformation, promotes finer bainitic features and increases the resistance to plastic deformation.


    \item The comparison of bainitic ferrite width between phase-field simulations and experiments demonstrates that the model captures the correct temperature-dependent trend. Both simulation and experiment show an increase in bainitic ferrite width with increasing holding temperature from \(673~\mathrm{K}\) to \(723~\mathrm{K}\). However, the simulations overestimates the absolute bainitic ferrite width, particularly at \(673~\mathrm{K}\). This indicates qualitative agreement in morphology evolution, while further calibration is required to achieve quantitative accuracy.


    \item The post-heat-treatment tensile simulations reproduce the experimentally observed dependence of the mechanical response on holding temperature. For both holding temperatures, the simulations performed with $r=0.25~\mathrm{s^{-1}}$ were calibrated against the experimental tensile curves and show close agreement with the measured response. The condition transformed at $673~\mathrm{K}$ exhibits a higher flow stress than that at $723~\mathrm{K}$, consistent with the finer bainitic morphology and the higher transformation driving force associated with the lower holding temperature. In comparison, the simulations performed with $r=0.5~\mathrm{s^{-1}}$ predict higher strength levels, which can be attributed to the stronger cooling condition, accelerated transformation kinetics, enhanced microstructural refinement, and higher transformation-induced internal stresses.

    \item The Barlat91 phenomenological yield surfaces reveal a clear anisotropic plastic response of the bainitic microstructures. The non-circular yield loci demonstrate that the mechanical response cannot be adequately described using an isotropic von Mises yield criterion. The largest yield surface is obtained for the \(673~\mathrm{K}\), \(r=0.5~\mathrm{s^{-1}}\) condition, indicating the highest predicted yield resistance. This response results from the combined effects of lower transformation temperature, faster heat extraction, bainitic ferrite refinement and retained internal stresses.

    \item The yield-surface analysis further shows that the effect of holding temperature is loading-path dependent. Under uniaxial loading, the response is more sensitive to the orientation of bainitic ferrite features relative to the loading direction. In the shear-dominated regions of the yield locus, the difference between the \(673~\mathrm{K}\) and \(723~\mathrm{K}\) cases becomes less pronounced, suggesting that shear-like deformation samples a broader part of the microstructural constraint network and partially averages the influence of individual bainitic features.

    \item The Barlat91 yield function provides an efficient phenomenological representation of anisotropic yielding and offers a useful bridge between microstructure-resolved phase-field/crystal-plasticity simulations and continuum-scale plasticity modelling. However, since only the \(x\)-direction tensile response was experimentally validated, the remaining loading directions should be interpreted as model-based extrapolations. 

    \item Overall, the results establish a consistent process-microstructure-property relationship for bainitic steels heat treatment. The heat-treatment parameters control the transformation kinetics, which determine bainitic morphology, internal-stress development, tensile strength and anisotropic yielding behaviour. The study demonstrates that accurate prediction of mechanical properties requires simultaneous consideration of phase fraction, bainitic ferrite morphology, transformation-induced internal stress and loading-direction-dependent plasticity.
    
\end{itemize}

\textbf{Data Availability} \par 
The phase-field simulation data used in this study can be shared upon reasonable request to the authors.

\textbf{Acknowledgements} \par 
The authors acknowledge funding by the  Federal Ministry of Research, Technology and Space (BMFTR), Germany  - project 13XP5226E (DiStEL).

\bibliographystyle{elsarticle-num}

\bibliography{references}

@article{gulbay2023influence,
  title={Influence of transformation temperature on the high-cycle fatigue performance of carbide-bearing and carbide-free bainite},
  author={Gulbay, Oguz and Ackermann, Marc and Gramlich, Alexander and Durmaz, Ali Riza and Steinbach, Ingo and Krupp, Ulrich},
  journal={steel research international},
  volume={94},
  number={12},
  pages={2300238},
  year={2023},
  publisher={Wiley Online Library}
}

@article{nerella2025automated,
  title={Automated Workflow for Phase-Field Simulations: Unveiling the Impact of Heat-Treatment Parameters on Bainitic Microstructure in Steel},
  author={Nerella, Dhanunjaya K and Ali, Muhammad Adil and Salama, Hesham and Gulbay, Oguz and Ackermann, Marc and Shchyglo, Oleg and Krupp, Ulrich and Steinbach, Ingo},
  journal={Advanced Engineering Materials},
  volume={27},
  number={8},
  pages={2400905},
  year={2025},
  publisher={Wiley Online Library}
}

@article{bakhtiari2009effect,
  title={The effect of bainite morphology on the mechanical properties of a high bainite dual phase (HBDP) steel},
  author={Bakhtiari, R and Ekrami, A},
  journal={Materials Science and Engineering: A},
  volume={525},
  number={1-2},
  pages={159--165},
  year={2009},
  publisher={Elsevier}
}

@article{zhao2019effect,
  title={Effect of bainite morphology on deformation compatibility of mesostructure in ferrite/bainite dual-phase steel: Mesostructure-based finite element analysis},
  author={Zhao, Zhong-tao and Wang, Xue-song and Qiao, Gui-ying and Zhang, Shi-yu and Liao, Bo and Xiao, Fu-ren},
  journal={Materials \& Design},
  volume={180},
  pages={107870},
  year={2019},
  publisher={Elsevier}
}

@article{ohtani1990morphology,
  title={Morphology and properties of low-carbon bainite},
  author={Ohtani, H and Okaguchi, S and Fujishiro, Y and Ohmori, Y},
  journal={Metallurgical transactions A},
  volume={21},
  number={3},
  pages={877--888},
  year={1990},
  publisher={Springer}
}

@article{steinbach1999generalized,
  title={A generalized field method for multiphase transformations using interface fields},
  author={Steinbach, Ingo and Pezzolla, F},
  journal={Physica D: Nonlinear Phenomena},
  volume={134},
  number={4},
  pages={385--393},
  year={1999},
  publisher={Elsevier}
}

@article{shchyglo2019phase,
  title={Phase-field simulation of martensite microstructure in low-carbon steel},
  author={Shchyglo, Oleg and Du, Guanxing and Engels, Jenni K and Steinbach, Ingo},
  journal={Acta Materialia},
  volume={175},
  pages={415--425},
  year={2019},
  publisher={Elsevier}
}

@article{steinbach2009phase,
  title={Phase-field models in materials science},
  author={Steinbach, Ingo},
  journal={Modelling and simulation in materials science and engineering},
  volume={17},
  number={7},
  pages={073001},
  year={2009}
}

@article{heo2014phase,
  title={Phase-field modeling of displacive phase transformations in elastically anisotropic and inhomogeneous polycrystals},
  author={Heo, Tae Wook and Chen, Long-Qing},
  journal={Acta Materialia},
  volume={76},
  pages={68--81},
  year={2014},
  publisher={Elsevier}
}

@article{moelans2008introduction,
  title={An introduction to phase-field modeling of microstructure evolution},
  author={Moelans, Nele and Blanpain, Bart and Wollants, Patrick},
  journal={Calphad},
  volume={32},
  number={2},
  pages={268--294},
  year={2008},
  publisher={Elsevier}
}

@article{levitas2013phase,
  title={Phase-field theory for martensitic phase transformations at large strains},
  author={Levitas, Valery I},
  journal={International Journal of Plasticity},
  volume={49},
  pages={85--118},
  year={2013},
  publisher={Elsevier}
}

@article{miehe2002strain,
  title={Strain-driven homogenization of inelastic microstructures and composites based on an incremental variational formulation},
  author={Miehe, Christian},
  journal={International Journal for numerical methods in engineering},
  volume={55},
  number={11},
  pages={1285--1322},
  year={2002},
  publisher={Wiley Online Library}
}

@article{roters2010overview,
  title={Overview of constitutive laws, kinematics, homogenization and multiscale methods in crystal plasticity finite-element modeling: Theory, experiments, applications},
  author={Roters, Franz and Eisenlohr, Philip and Hantcherli, Luc and Tjahjanto, Denny Dharmawan and Bieler, Thomas R and Raabe, Dierk},
  journal={Acta materialia},
  volume={58},
  number={4},
  pages={1152--1211},
  year={2010},
  publisher={Elsevier}
}

@article{kocks1975thermodynamics,
  title={Thermodynamics and kinetics of slip},
  author={Kocks, Ulrich Fred and As, Argon and Mf, Ashby},
  year={1975}
}

@article{shchyglo2024efficient,
  title={Efficient finite strain elasticity solver for phase-field simulations},
  author={Shchyglo, Oleg and Ali, Muhammad Adil and Salama, Hesham},
  journal={NPJ Computational Materials},
  volume={10},
  number={1},
  pages={52},
  year={2024},
  publisher={Nature Publishing Group UK London}
}

@article{molnar1969newton,
  title={Newton's thermometer: a model for testing Newton's law of cooling.},
  author={Molnar, GEORGE W},
  journal={The Physiologist},
  volume={12},
  number={1},
  pages={9--20},
  year={1969}
}

@article{hajizad2019influence,
  title={Influence of microstructure on mechanical properties of bainitic steels in railway applications},
  author={Hajizad, Omid and Kumar, Ankit and Li, Zili and Petrov, Roumen H and Sietsma, Jilt and Dollevoet, Rolf},
  journal={Metals},
  volume={9},
  number={7},
  pages={778},
  year={2019},
  publisher={MDPI}
}

@article{fan2022effect,
  title={Effect of microstructure on wear and rolling contact fatigue behaviors of bainitic/martensitic rail steels},
  author={Fan, Yusong and Gui, Xiaolu and Liu, Miao and Wang, Xi and Bai, Bingzhe and Gao, Guhui},
  journal={Wear},
  volume={508},
  pages={204474},
  year={2022},
  publisher={Elsevier}
}

@article{fielding2013bainite,
  title={The bainite controversy},
  author={Fielding, LCD},
  journal={Materials Science and Technology},
  volume={29},
  number={4},
  pages={383--399},
  year={2013},
  publisher={SAGE Publications Sage UK: London, England}
}

@article{steinbach2024highly,
  title={Highly complex materials processes as understood by phase-field simulations: Additive manufacturing, bainitic transformation in steel and high-temperature creep of superalloys},
  author={Steinbach, Ingo and Uddagiri, Murali and Salama, Hesham and Ali, Muhammad Adil and Shchyglo, Oleg},
  journal={MRS Bulletin},
  volume={49},
  number={6},
  pages={583--593},
  year={2024},
  publisher={Springer}
}

@article{steinbach2013phase,
  title={Phase-field model for microstructure evolution at the mesoscopic scale},
  author={Steinbach, Ingo},
  journal={Annual Review of Materials Research},
  volume={43},
  number={1},
  pages={89--107},
  year={2013},
  publisher={Annual Reviews}
}

@article{barlat91,
  title={A six-component yield function for anisotropic materials},
  author={Barlat, Fr{\'e}d{\'e}ric and Lege, Daniel J and Brem, John C},
  journal={International journal of plasticity},
  volume={7},
  number={7},
  pages={693--712},
  year={1991},
  publisher={Elsevier}
}

@article{khalfallah2015influence,
  title={Influence of the characteristics of the experimental data set used to identify anisotropy parameters},
  author={Khalfallah, Ali and Alves, Jos{\'e} Lu{\'\i}s and Oliveira, Marta Cristina and Menezes, Lu{\'\i}s Filipe},
  journal={Simulation Modelling Practice and Theory},
  volume={53},
  pages={15--44},
  year={2015},
  publisher={Elsevier}
}

@article{chaparro2008material,
  title={Material parameters identification: Gradient-based, genetic and hybrid optimization algorithms},
  author={Chaparro, Bruno M and Thuillier, Sandrine and Menezes, Lu{\i}s Filipe and Manach, Pierre-Yves and Fernandes, Jose Valdemar},
  journal={Computational Materials Science},
  volume={44},
  number={2},
  pages={339--346},
  year={2008},
  publisher={Elsevier}
}

@article{zhao2025review,
  title={Review on Heat Treatment and Surface Modification Technology of High-Strength Bainite Steels},
  author={Zhao, Siyang and Liu, Man and Tian, Junyu and Dai, Fangqin and Xu, Guang},
  journal={steel research international},
  volume={96},
  number={7},
  pages={2400685},
  year={2025},
  publisher={Wiley Online Library}
}

@article{kaikkonen2023evaluation,
  title={Evaluation of a processing route and microstructural characteristics for the development of ultrafine bainite in low-temperature ausformed medium-carbon steels},
  author={Kaikkonen, Pentti and Somani, Mahesh C and Pohjonen, Aarne and Javaheri, Vahid and K{\"o}mi, Jukka},
  journal={Journal of Materials Engineering and Performance},
  volume={32},
  number={17},
  pages={7846--7857},
  year={2023},
  publisher={Springer}
}

@book{soliman2008phase,
  title={Phase transformations and mechanical properties of new austenite-stabilised bainite steels},
  author={Soliman, Mohamed A and others},
  year={2008},
  publisher={Pieper}
}

@article{xiao2024design,
  title={Design of cooling route for carbide-free bainitic rail steels and resultant microstructures and properties},
  author={Xiao, Naiyou and Fei, Junjie and Li, Meiying and Zhou, Jianhua and Jia, Tao},
  journal={Materials Science and Engineering: A},
  volume={891},
  pages={145936},
  year={2024},
  publisher={Elsevier}
}

@article{nanda2019third,
  title={Third generation of advanced high-strength steels: Processing routes and properties},
  author={Nanda, Tarun and Singh, Vishal and Singh, Virender and Chakraborty, Arnab and Sharma, Sandeep},
  journal={Proceedings of the Institution of Mechanical Engineers, Part L: Journal of Materials: Design and Applications},
  volume={233},
  number={2},
  pages={209--238},
  year={2019},
  publisher={SAGE Publications Sage UK: London, England}
}

@article{wang2022novel,
  title={A novel route to improve the fatigue properties of aviation M50 steel via tailoring the bainite content and cold deformation},
  author={Wang, Feng and Du, Yuchen and Qian, Dongsheng and Cao, Nana and Hua, Lin and Wu, Min},
  journal={Journal of Materials Research and Technology},
  volume={18},
  pages={3857--3871},
  year={2022},
  publisher={Elsevier}
}

@article{liang2014study,
  title={A study of the influence of thermomechanical controlled processing on the microstructure of bainite in high strength plate steel},
  author={Liang, Xiaojun and DeArdo, Anthony J},
  journal={Metallurgical and Materials Transactions A},
  volume={45},
  number={11},
  pages={5173--5184},
  year={2014},
  publisher={Springer}
}

@article{steinbach2006multi,
  title={Multi phase field model for solid state transformation with elastic strain},
  author={Steinbach, Ingo and Apel, Markus},
  journal={Physica D: Nonlinear Phenomena},
  volume={217},
  number={2},
  pages={153--160},
  year={2006},
  publisher={Elsevier}
}

@article{Peirce1983,
  title={Material rate dependence and localized deformation in crystalline solids},
  author={Peirce, Daniel and Asaro, Robert J and Needleman, A},
  journal={Acta metallurgica},
  volume={31},
  number={12},
  pages={1951--1976},
  year={1983},
  publisher={Elsevier}
}

@article{nerella20262d,
  title={A 2D and 3D microstructural descriptor approach to bainitic steels: Insights from phase-field simulations},
  author={Nerella, Dhanunjaya K and Jogi, Tushar and Gulbay, Oguz and Ali, Muhammad Adil and Shchyglo, Oleg and Steinbach, Ingo},
  journal={Acta Materialia},
  pages={122340},
  year={2026},
  publisher={Elsevier}
}

@article{Zhang2016,
  title={A virtual laboratory using high resolution crystal plasticity simulations to determine the initial yield surface for sheet metal forming operations},
  author={Zhang, Haiming and Diehl, Martin and Roters, Franz and Raabe, Dierk},
  journal={International Journal of Plasticity},
  volume={80},
  pages={111--138},
  year={2016},
  publisher={Elsevier}
}

@incollection{Banabic2010,
  title={Plastic behaviour of sheet metal},
  author={Banabic, Dorel},
  booktitle={Sheet metal forming processes},
  pages={27--140},
  year={2010},
  publisher={Springer}
}

@article{Shchyglo2024,
  title = {Efficient finite strain elasticity solver for phase-field simulations},
  volume = {10},
  ISSN = {2057-3960},
  url = {http://dx.doi.org/10.1038/s41524-024-01235-4},
  DOI = {10.1038/s41524-024-01235-4},
  number = {1},
  journal = {npj Computational Materials},
  publisher = {Springer Science and Business Media LLC},
  author = {Shchyglo,  Oleg and Ali,  Muhammad Adil and Salama,  Hesham},
  year = {2024},
  month = Mar 
}

@article{Salama2024,
  title = {Phase-field simulation framework for modeling martensite and bainite formation in steel},
  volume = {241},
  ISSN = {0927-0256},
  url = {http://dx.doi.org/10.1016/j.commatsci.2024.113033},
  DOI = {10.1016/j.commatsci.2024.113033},
  journal = {Computational Materials Science},
  publisher = {Elsevier BV},
  author = {Salama,  Hesham and Ali,  Muhammad Adil and Shchyglo,  Oleg and Steinbach,  Ingo},
  year = {2024},
  month = May,
  pages = {113033}
}

@article{Ali2020,
  title = {Role of coherency loss on rafting behavior of Ni-based superalloys},
  volume = {171},
  ISSN = {0927-0256},
  url = {http://dx.doi.org/10.1016/j.commatsci.2019.109279},
  DOI = {10.1016/j.commatsci.2019.109279},
  journal = {Computational Materials Science},
  publisher = {Elsevier BV},
  author = {Ali,  Muhammad Adil and G\"{o}rler,  Johannes V. and Steinbach,  Ingo},
  year = {2020},
  month = Jan,
  pages = {109279}
}

@article{Yeddu2012,
  title = {Three-dimensional phase-field modeling of martensitic microstructure evolution in steels},
  volume = {60},
  ISSN = {1359-6454},
  url = {http://dx.doi.org/10.1016/j.actamat.2011.11.039},
  DOI = {10.1016/j.actamat.2011.11.039},
  number = {4},
  journal = {Acta Materialia},
  publisher = {Elsevier BV},
  author = {Yeddu,  Hemantha Kumar and Malik,  Amer and Ågren,  John and Amberg,  Gustav and Borgenstam,  Annika},
  year = {2012},
  month = Feb,
  pages = {1538–1547}
}

@article{Levitas2013,
  title = {Phase-field theory for martensitic phase transformations at large strains},
  volume = {49},
  ISSN = {0749-6419},
  url = {http://dx.doi.org/10.1016/j.ijplas.2013.03.002},
  DOI = {10.1016/j.ijplas.2013.03.002},
  journal = {International Journal of Plasticity},
  publisher = {Elsevier BV},
  author = {Levitas,  Valery I.},
  year = {2013},
  month = Oct,
  pages = {85–118}
}

@article{Levitas2015,
  title = {Interaction between phase transformations and dislocations at the nanoscale. Part 1. General phase field approach},
  volume = {82},
  ISSN = {0022-5096},
  url = {http://dx.doi.org/10.1016/j.jmps.2015.05.005},
  DOI = {10.1016/j.jmps.2015.05.005},
  journal = {Journal of the Mechanics and Physics of Solids},
  publisher = {Elsevier BV},
  author = {Levitas,  Valery I. and Javanbakht,  Mahdi},
  year = {2015},
  month = Sept,
  pages = {287–319}
}

@article{Javanbakht2015,
  title = {Interaction between phase transformations and dislocations at the nanoscale. Part 2: Phase field simulation examples},
  volume = {82},
  ISSN = {0022-5096},
  url = {http://dx.doi.org/10.1016/j.jmps.2015.05.006},
  DOI = {10.1016/j.jmps.2015.05.006},
  journal = {Journal of the Mechanics and Physics of Solids},
  publisher = {Elsevier BV},
  author = {Javanbakht,  Mahdi and Levitas,  Valery I.},
  year = {2015},
  month = Sept,
  pages = {164–185}
}

\end{document}